\documentclass[twocolumn,showpacs,amsmath,amssymb,aps,pre,reprint,longbibliography]{revtex4-1}
\usepackage{graphicx}
\begin{document}

\title{
Active Matter Shepherding and Clustering in Inhomogeneous Environments   
} 
\author{
P. Forg{\' a}cs$^{1}$, A. Lib{\' a}l$^{1}$, C. Reichhardt$^{2}$, and C. J. O. Reichhardt$^{2}$ 
} 
\affiliation{
$^{1}$ Mathematics and Computer Science Department, Babe{\c s}-Bolya University, Cluj 400084, Romania\\
$^{2}$ Theoretical Division and Center for Nonlinear Studies,
Los Alamos National Laboratory, Los Alamos, New Mexico 87545, USA}

\date{\today}
\begin{abstract}
  We consider a mixture of active and passive run-and-tumble disks in an inhomogeneous environment where only half of the sample contains quenched disorder or pinning. The disks are initialized in a fully mixed state of uniform density. We identify several distinct dynamical phases as a function of motor force and pinning density. At high pinning densities and high motor forces, there is a two step process initiated by a rapid accumulation of both active and passive disks in the pinned region, which produces a large density gradient in the system. This is followed by a slower species phase separation process where the inactive disks are shepherded by the active disks into the pin-free region, forming a non-clustered fluid and producing a more uniform density with species phase separation. For higher pinning densities and low motor forces, the dynamics becomes very slow and the system maintains a strong density gradient. For weaker pinning and large motor forces, a floating clustered state appears and the time averaged density of the system is uniform. We illustrate the appearance of these phases in a dynamic phase diagram.
\end{abstract}
\maketitle

Active matter,
or
self-motile particles or agents,
arise in a variety of contexts including biological as well as designed systems 
such as self-propelled colloids and artificial swimmers
\cite{Marchetti13,Bechinger16,Gompper20}. Due to  
their intrinsic non-equilibrium nature,
such systems exhibit collective behaviors which are absent 
in systems with only thermal or Brownian motion.
One of the best known examples of such an effect is the
motility induced phase separation found for
a collection of interacting disks.
In the Brownian limit, at lower densities the disks form a uniform 
liquid phase; however, if the disks
are active, there can be a transition to
a self-clustered or phase separated state composed of
regions of high density or solid clusters
coexisting with a low density active gas
\cite{Bechinger16,Fily12,Redner13,Palacci13,Buttinoni13,Cates15}.
This phase separation can occur 
even when all the pairwise interactions between particles are repulsive. 

Active matter 
can also exhibit a variety of other effects when it is
coupled to complex environments \cite{Bechinger16}, 
such as clustering along walls
\cite{Fily14a,Sartori18,Speck20,Das20,Fazli21}, 
activity induced depletion forces
\cite{Ray14,Ni15,Leite16,Mallory18,Kjeldbjerg21},
ratchet behavior or spontaneous directed motion
through funnels or asymmetric shapes
\cite{Galajda07,Tailleur09,Ai16,Reichhardt17a,Borba20},
novel transport in 
maze geometries \cite{Khatami16,Yang18},
and avalanches or activity induced clogging phenomena in constricted
geometries \cite{Reichhardt18c,Caprini20,Shi20}.
Active matter systems
with random quenched disorder can undergo
activity induced jamming \cite{Reichhardt14},
trapping \cite{Zeitz17,Sandor17b,Morin17a},
non-monotonic mobility
\cite{Chepizhko13,Reichhardt14,Bertrand18,Chepizhko19,Bhattacharjee19,Breoni20},
and different types of motion under external drift forces
\cite{Sandor17a,Reichhardt18a,Morin17,Bijnens21,Chardac21}.

The quenched disorder can take the form
of obstacles which act as repulsive barriers
or pinning sites which behave as localized traps.
For dense pinning sites,
the activity induced clustering effect can be suppressed when the
active particles remain spread out due to trapping in the pinning sites
\cite{Sandor17b}.
In contrast, obstacles can promote 
activity induced clustering \cite{Reichhardt14} 
by acting as nucleation sites \cite{Reichhardt14a}.
The enhancement or reduction of activity included clustering
by random disorder depends on the
density and strength of the disorder.
In addition to random disorder,
there have been a variety of studies examining active matter
coupled to periodic
substrates which reveal directional locking effects
\cite{Volpe11,Reichhardt20,BrunCosmeBruny20}, 
nonlinear transport \cite{Pattanayak19,Schakenraad20,Ribeiro20,Reichhardt21a},
and commensuration effects \cite{Reinken20,Reichhardt21}. 

In non-active matter systems with quenched disorder, 
such as vortices in superconductors \cite{Marchetti99,Banerjee03},
colloidal particles \cite{Seshadri93,Nagamanasa16},
and skyrmions \cite{Reichhardt20aa},
it is known that the dynamics of the system can be modified strongly
if the quenched disorder is inhomogeneous.
An example of such disorder is a sample containing
extended regions of strong pinning 
coexisting with regions where the pinning is weak or absent. 
Here, under an applied drive or
increasing temperature, the system exhibits high mobility in the unpinned
regions and reduced mobility in the pinned regions
\cite{Marchetti99,Seshadri93,Reichhardt20aa}.
Also,
as the temperature is decreased from a high value,
a glass or solid state first begins to form in the pinned region,
with glassy 
behavior gradually spreading into the non-pinned regions
\cite{Banerjee03,Seshadri93}. 
In inhomogeneous pinning environments, application of an external drive
can produce an accumulation of particles 
along the interface between the pinned and pin-free
regions \cite{Reichhardt20aa}.
These behaviors suggest that active matter
moving in a sample containing coexisting pinned and non-pinned regions
should also show significantly different behavior
from active systems in uniform pinning.   
  
In this work, we numerically examine a bidisperse assembly of
run-and-tumble disks interacting with a substrate that is populated on
only one half by
quenched disorder in the form of pinning sites or obstacles.
Half of the
disks are active while the other half are passive.
We vary the number of pinning sites and the
active motor force, and initialize the system in a fully
mixed state with a uniform density across the sample. 
For high motor forces and low pinning densities,
the system forms an active mobile cluster phase in which the clusters 
contain both active and passive disks. These
clusters spend equal amounts of time
in the pinned and non-pinned regions, and the clusters act as 
large scale objects which overcome the pinning effects
to form what we call a floating cluster state
with a uniform time-averaged density.
At high motor forces and high pinning densities, we
observe a two step dynamical process involving a transition from the uniform
state to a density phase separated state followed by a coarsening process in
which the different disk species segregate.
In the first stage,
a large build up of both species of
disks into low mobility clusters occurs in the pinned region,
producing a large density gradient
between
the pinned and non-pinned regions.
The second stage
is a
slower species phase separation
process in which active disks gradually shepherd the 
non-active disks out of the
pinned regions,
producing a phase separated state with a more uniform density
in the long time limit.
For other parameters,
clusters can start to form along the edge of the pinned region, followed
by the slow motion of a well defined
density front
into the pinned region.
This process becomes slower as the motor force decreases.
For the case of obstacles
instead of pinning sites, we observe similar dynamics; however,
the floating clustered state is lost
and the shepherding behavior appears at much lower obstacle densities.
When all the disks are active and the pinning is strong,
the system only forms a low mobility clustered state
in the pinned region and maintains a low density of active
disks in the unpinned region.      

We note that Chepizhko and Peruani \cite{Chepizhko13a}
have also considered individual active particles interacting
with quenched disorder in the form of obstacles placed in only
half of the sample.
They found that for sufficiently high
obstacle densities, the active particle could become locally trapped.
This model differs significantly from the present work, where we focus
on strong collective interactions between
the active particles.

\section{Simulation}

\begin{figure}
\includegraphics[width=\columnwidth]{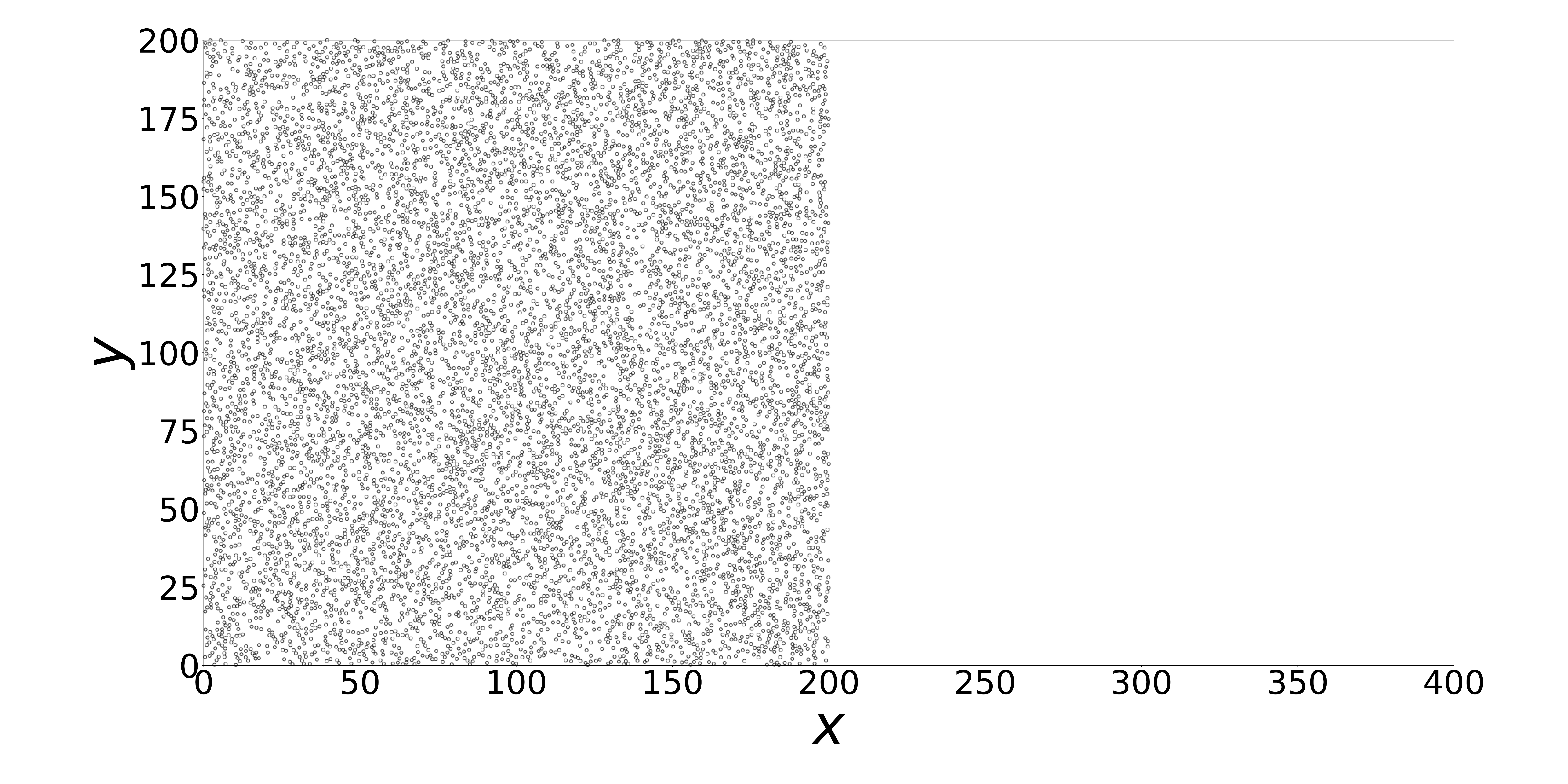}
\caption{
Image of the substrate showing that the right half of the sample is empty and
the left half contains $N_p$ nonoverlapping pinning sites (dots).
}
\label{fig:1}
\end{figure}

We consider a two-dimensional system of size $L_x=400$ and $L_y=200$
with
periodic boundary conditions in the $x$ and $y$-directions.
As illustrated in Fig.~\ref{fig:1}, the system is divided into two regions.
The right region has no substrate,
while the left region contains $N_{p}$ nonoverlapping pinning
sites that are modeled as attractive wells.
We initialize the system with a random uniform distribution
of $N_{d}$ disks with radius $r_{a}$.
The disk-disk interactions are represented by a stiff
harmonic repulsion and the disk density is   
$\phi = N_{d}\pi r^2_{a}/L_{x}L_{y}$.
The disks are divided into two populations, A and B,
where $N_{A}$ of the disks are active and experience self propulsion, while
the remaining $N_B=N_d-N_A$ disks are passive and can move only in response to
other disks or the pinning sites.
The equation of motion of disk $i$ is
\begin{equation} 
\alpha_d {\bf v}_{i}  =
{\bf F}^{dd}_{i} + {\bf F}^{m}_{i} + {\bf F}^{obs}_{i} \ .
\end{equation}
The disk velocity is ${\bf v}_{i} = {d {\bf r}_{i}}/{dt}$,  
where ${\bf r}_{i}$ is the disk position
and the damping constant $\alpha_d = 1.0$.
The disk-disk force is given by the harmonic repulsive potential 
${\bf F}_{dd} = \sum_{i\neq j}^{N_d}k(2r_{a} - |{\bf r}_{ij}|)\Theta(2r_{a} - |{\bf r}_{ij}|) {\hat {\bf r}_{ij}}$,
where $\Theta$ is the Heaviside step function,
${\bf r}_{ij} = {\bf r}_{i} - {\bf r}_{j}$, and
$\hat {\bf r}_{ij}  = {\bf r}_{ij}/|{\bf r}_{ij}|$. We set
$k = 20$ and $r_{a} = 1.0$ for both species $A$ and $B$. 
The active disks obey run-and-tumble dynamics in which
a motor force $F_{M}$ is exerted on the disk in a randomly chosen direction
during a run time of $\tau_{l}$ before instantaneously changing to a new
randomly chosen direction during the next run time.
We define the run length
$l_{r}$ as the distance a disk would
travel in the absence of pinning sites or disk-disk collisions,
$l_{r} = F_{M}\tau_{l}\delta t$,
where $\delta t=0.005$ is the simulation time step.
We vary $\tau_{l}$ over the range $\tau_l = 40,000$ to $80,000$ time steps
and the motor force $F_M$ over the range
$F_M=0.175$ to $2.0$.
For the non-active disks, the equation of motion
is the same except $F_{M} = 0$.  
We use a total simulation time of
$2\times 10^7$.
We fix the total number of disks to give a system density of
$\phi = 0.55$. 
The pinning sites are modeled
as harmonic traps with
strength $F_{p} = 2.5$ and radius $r_p=0.5$, such that a single pinning
site can capture at most one disk.
After initialization, we measure the time evolution
of the local disk density $\phi_l$ as a function of $x$ averaged over $y$
for all disks ($\phi_l^t$), only the active disks ($\phi_l^a$), and
only the passive disks ($\phi_l^p$).
We also measure the ratio of the average disk density in the unpinned,
$\phi_u$, and pinned, $\phi_p$, portions of the sample,
giving the density ratio of all disks $R_t=\phi_u^t/\phi_p^t$,
of the active disks $R_a=\phi_u^a/\phi_p^a$, and
of the passive disks $R_p=\phi_u^p/\phi_p^p$.

\section{Results}

\begin{figure}
\includegraphics[width=0.9\columnwidth]{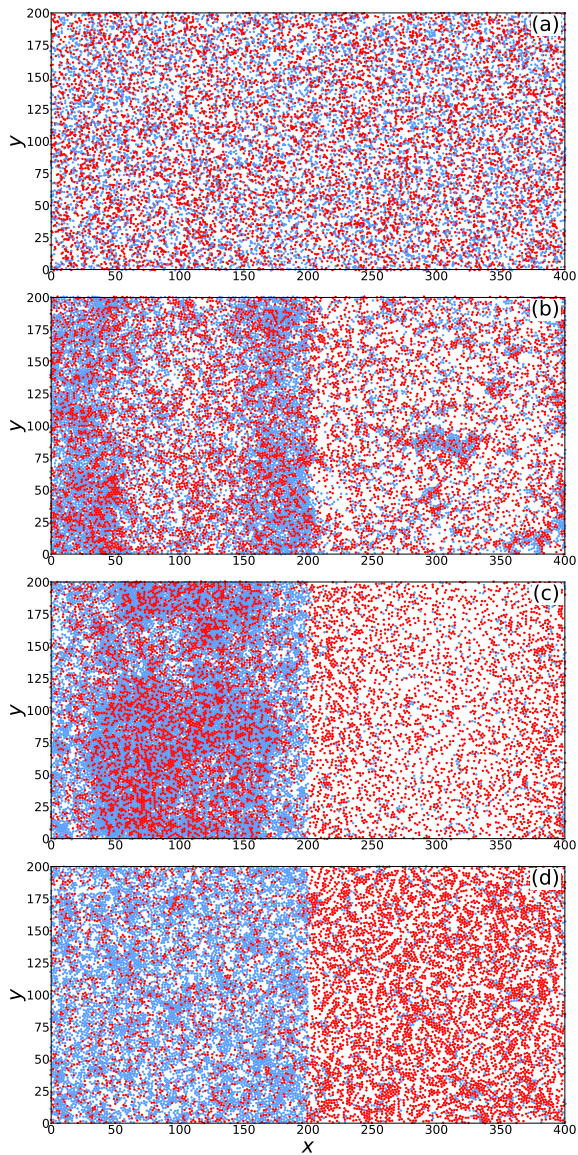}
\caption{
Images of the active disks (blue) and passive disks (red)    
for a system
with $F_{M} = 1.5$, $N_{p} = 9000$, and $l_{r} = 600$.
(a) The initial configuration in which the disk species are well mixed
and have a uniform density.
(b) At time $t = 3.9\times 10^4$,
a dense front of mixed species moves into the pinned region from both sides.
(c) At $t = 5\times 10^5$, there is a strong density imbalance and
the unpinned region contains a low density of mostly passive disks while
the pinned region contains a dense mixed state.
(d) At $t=2\times 10^7$,
the phase separation is more pronounced but the density difference between the
pinned and unpinned regions is diminished.
}
\label{fig:2}
\end{figure}

In Fig.~\ref{fig:2}
we plot the positions of the active
and passive disks
for a system 
with $F_{M} = 1.5$, $N_{p} = 9000$, and  $l_{r} = 600$.
The initial disk configuration
in Fig.~\ref{fig:2}(a)
has a uniform density with a uniform distribution of both disk species.
At time $t=3.9\times 10^4$ in Fig.~\ref{fig:2}(b),
a large density build up of both active and passive disks
appears at the pinning interface and begins to move into the pinned region.
The active and passive disks form
patches of denser clusters with six-fold ordering in the pinned area,
while in the non-pinned region 
a more uniform fluid like state appears. 
In Fig.~\ref{fig:2}(c) at $t = 5\times 10^5$, 
there is a strong density buildup in the pinned region while
the non-pinned region contains a small density of mostly passive disks,
with a density ratio of close to $2:1$ in the pinned and unpinned regions.
At longer times, a coarsening process occurs in which
the passive disks
are gradually shepherded out of the pinned region
by the active disks.
At $t= 2\times 10^7$ in Fig.~\ref{fig:2}(d),
the majority of the active disks have collected in
the pinned region and show pronounced clustering,
while the unpinned region
contains mostly passive disks being pushed around by a small
number of active disks.
The overall density of the system is more uniform compared to
Fig.~\ref{fig:2}(c).
The clusters in the pinned region generally have low mobility,
while weaker clustering of the passive disks in the
unpinned region occurs when the active disks push the
passive disks together.  

\begin{figure}
\includegraphics[width=\columnwidth]{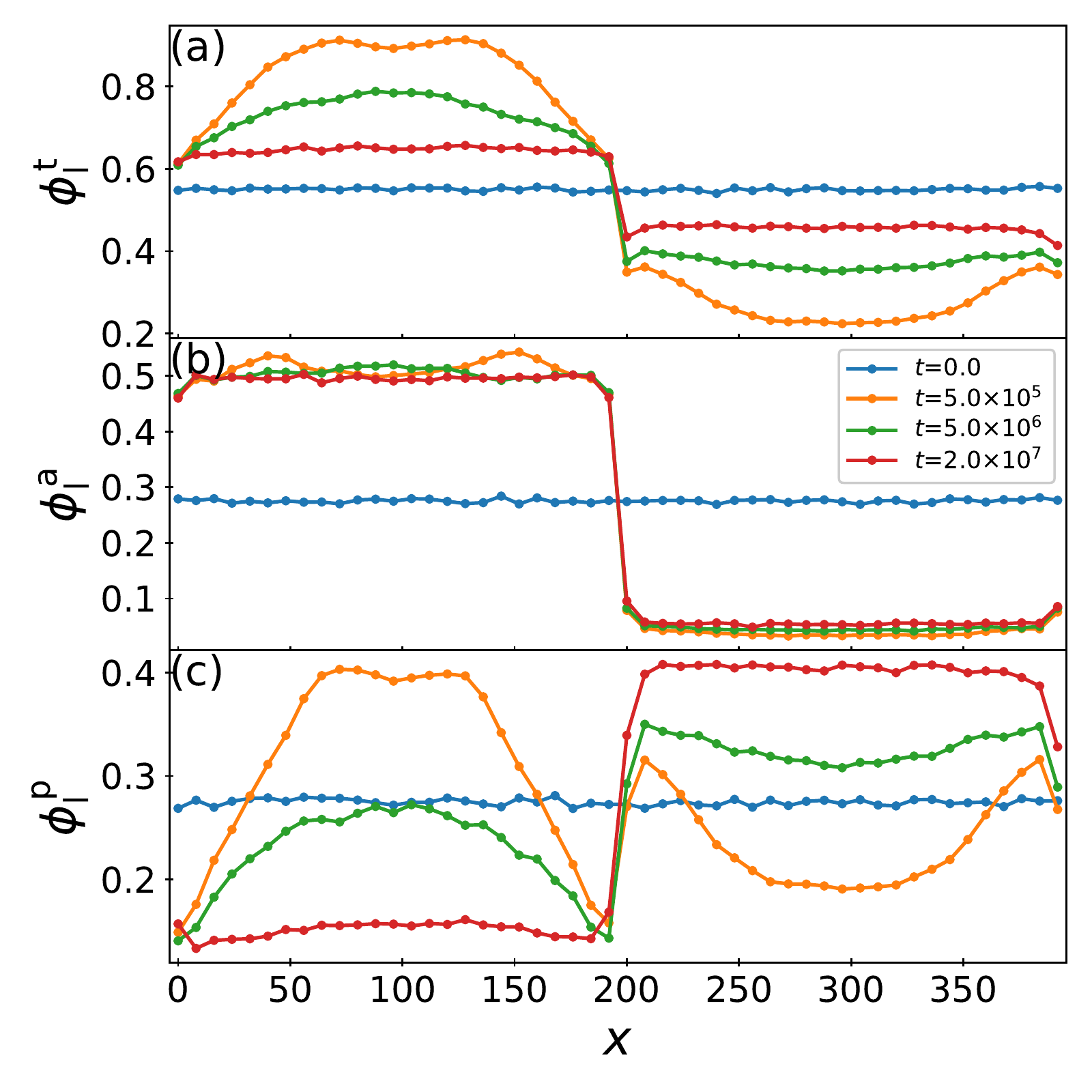}
\caption{Local densities $\phi_l$
as a function of $x$ for the sample in Fig.~\ref{fig:2}
with $F_M=1.5$, $N_p=9000$, and $l_r=600$ at
times $t = 0.0$ (blue) where all the densities are uniform,
$t = 5\times 10^5$ (orange),
$t = 5\times 10^6$ (green),
and $t = 2\times 10^7$ (red).  
(a) The local density of all disks $\phi_l^t$ versus $x$.
(b) The local density of active disks $\phi_l^a$ versus $x$.
(c) The local density of passive disks $\phi_l^p$ versus $x$.
}
\label{fig:3}
\end{figure}

To get a better picture of the time evolution,
in Fig.~\ref{fig:3}(a) we plot the local density
of all disks $\phi_l^t$
versus $x$ for the system in
Fig.~\ref{fig:2} at three different times, while
in Fig.~\ref{fig:3}(b,c) we show the local densities $\phi_l^a$ and $\phi_l^p$
of the active and passive disks, respectively,
versus $x$.
Pinning is present in the region with $x < 200$.
At $t = 0$, the density is uniform throughout the system.
For $t = 5\times 10^5$,
$\phi_l^t$ increases in the pinned region until the ratio of densities in the
pinned and unpinned regions is $2:1$.
The local density of active disks $\phi_l^a$
has a much stronger increase 
in the pinned region, with a ratio of close to $15:1$ for the density of
active particles in the pinned and unpinned regions.
At $t = 5\times 10^6$ and $t=2\times 10^7$,
$\phi_l^a$
in the pinned region remains fixed
while the local density of passive disks
$\phi_l^p$ in the pinned region drops.
This occurs when the active particles
shepherd the passive particles out of the pinned area,
and causes the local density of all disks
$\phi_l^t$ to become more uniform across the system.

\begin{figure}
\includegraphics[width=\columnwidth]{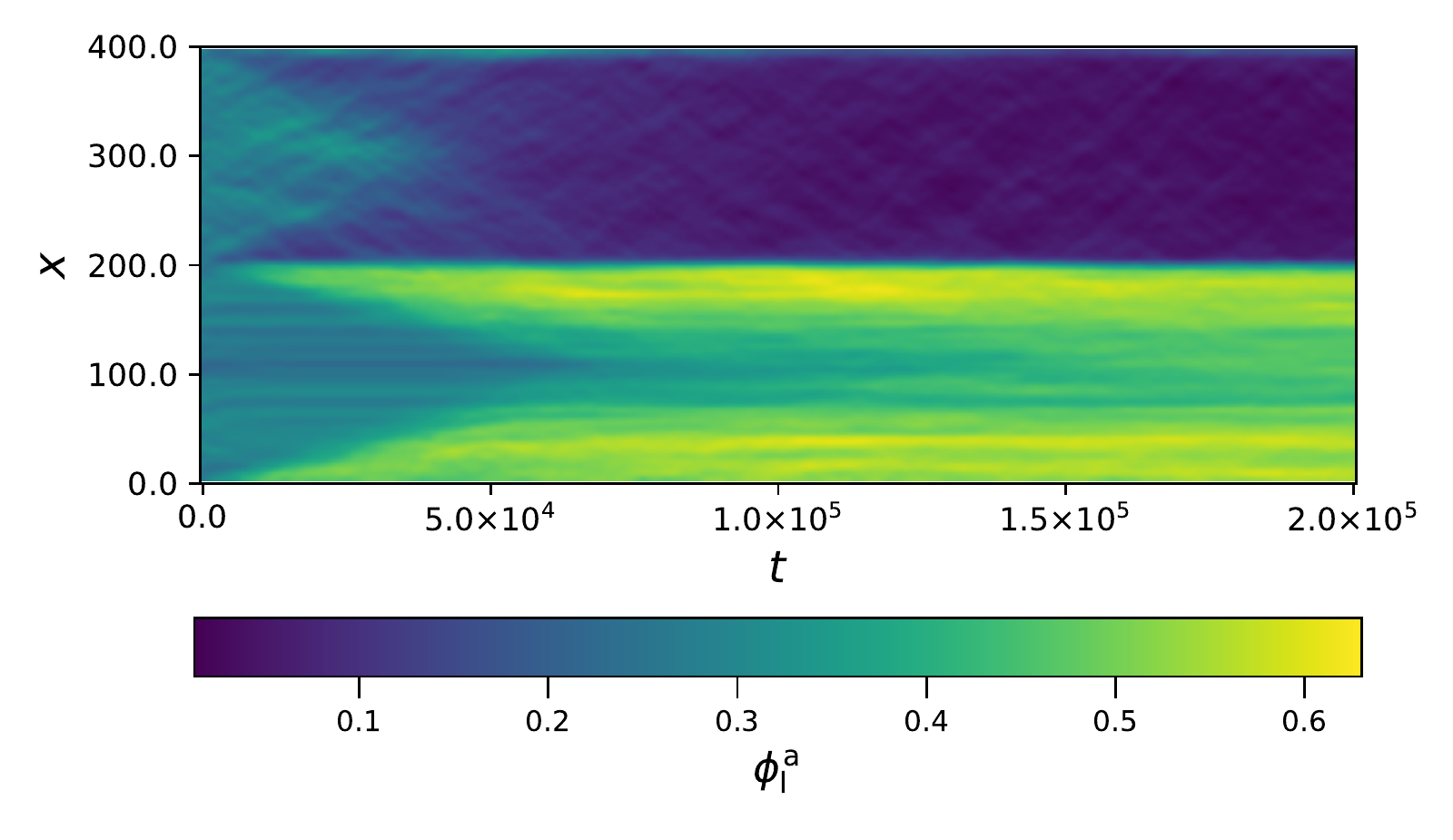}
\caption{Heat map of the local active disk density $\phi_l^a$ as a function
of position $x$ versus time
for the system in Fig.~\ref{fig:3} with $F_M=1.5$, $N_p=9000$,
and $l_r=600$.
The active disks move into the pinned regime, which is the region from
$x=0$ to $x=200$, in the form of a front of higher density.
 }
\label{fig:4}
\end{figure}

As shown in Fig.~\ref{fig:2},
the initial build up of active disks in the pinned region occurs
through the formation of a dense front that moves into the
pinned portion of the sample.
This is more clearly illustrated in Fig.~\ref{fig:4}
where we plot a heat map of $\phi_l^a$ as a function of $x$ position
versus time.
At $t=0$, $\phi_l^a$ is initially uniform,
and then local maxima develop at the
edges of the pinned region.
These high density areas become fronts of active disks
which move into the pinned region from either side
and gradually merge at the center of the pinned region over time.
For $t<5\times 10^4$, in the pinned region
there are bands of higher density which appear as horizontal
lines, indicating that the active disks have formed immobilized
dense clusters in these locations, while
in the unpinned region, the lines denoting high density areas
run at angles, indicating that the
active disks are in mobile clusters.

\begin{figure}
\includegraphics[width=\columnwidth]{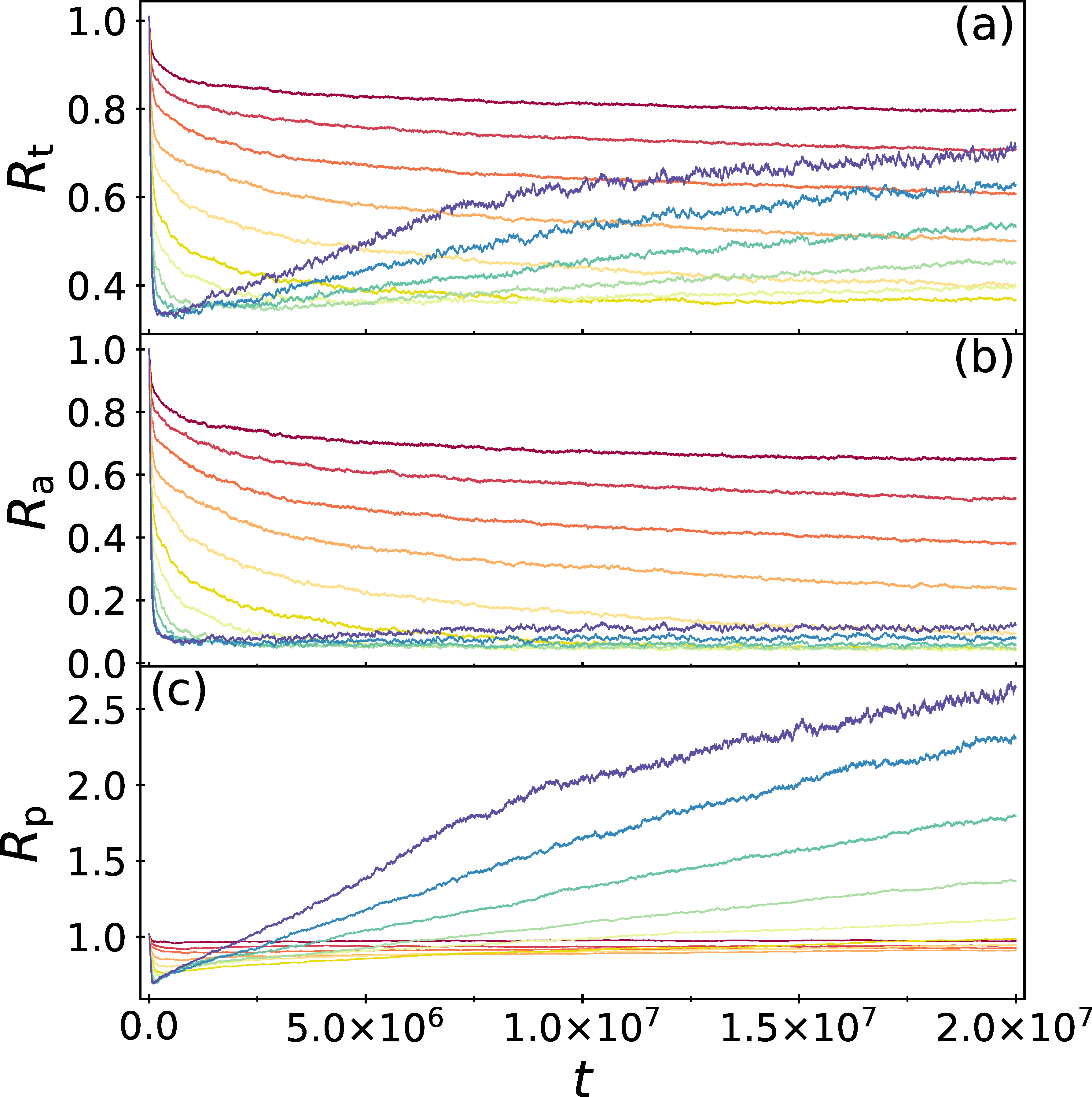}
\caption{Ratios $R$ of the average density $\phi_u$ in the unpinned region
to the average density $\phi_p$ in the
pinned region versus time for the system in Fig.~\ref{fig:3} with $N_p=9000$
and 
$l_r=600$ at varied
$F_M=0.5$, 0.6, 0.7, 0.8, 0.9, 1.0, 1.1, 1.2, 1.3, 1.4, and 1.5, from top left
to bottom left.
(a) The ratio $R_t=\phi_u^t/\phi_p^t$ of the density of all disks.
(b) The ratio $R_a=\phi_u^a/\phi_p^a$ of the active disk density.
(c) The ratio $R_p=\phi_u^p/\phi_p^p$ of the passive disk density.
We find a two stage evolution.
In the first stage, the active disks invade the pinned region,
while in the second stage,
the active particles shepherd the passive particles
out of the pinned region and into the unpinned region.
The duration of the first stage increases
as $F_{M}$ decreases.  
 }
\label{fig:5}
\end{figure}

\begin{figure}
\includegraphics[width=\columnwidth]{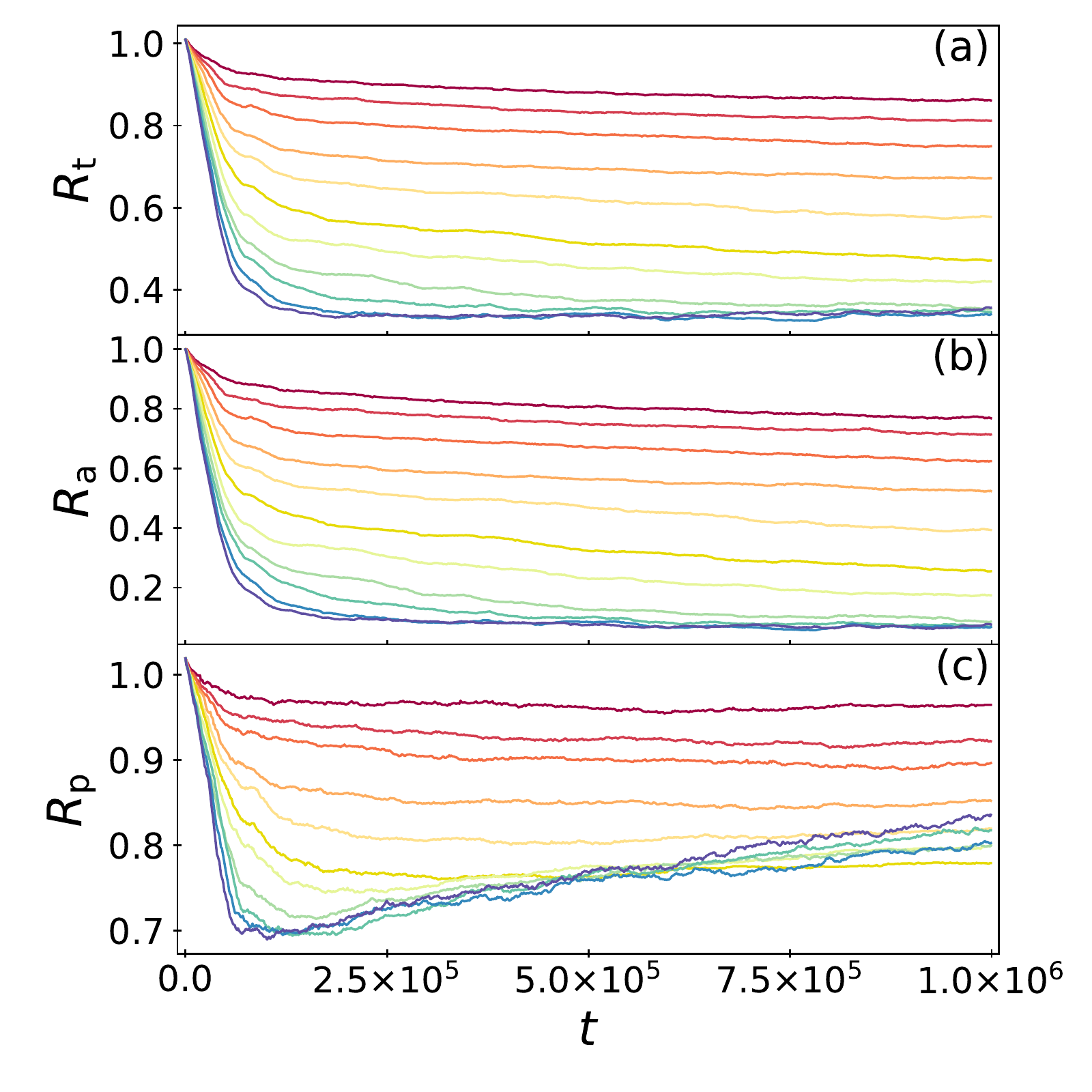}
\caption{A blow up of the
density ratio versus time curves from Fig.~\ref{fig:5} over the range
$t < 1.0\times 10^6$. Here $N_p=9000$, $l_r=600$, and 
$F_M=0.5$, 0.6, 0.7, 0.8, 0.9, 1.0, 1.1, 1.2, 1.3, 1.4, and 1.5, from top left
to bottom left.
(a) Density ratio for all disks $R_t=\phi_u^t/\phi_p^t$.
(b) Active disk density ratio $R_a=\phi_u^a/\phi_p^a$.
(c) Passive disk density ratio $R_p=\phi_u^p/\phi_p^p$.
The invasion of active disks
into the pinned region is accompanied by the dragging
of some of the passive disks
into the pinned
region.
This is followed by a crossover to passive
particle shepherding out of the pinned region, indicated by an increase
in $R_p$.
 }
\label{fig:6}
\end{figure}

In Fig.~\ref{fig:5}(a) we plot the ratio $R_t=\phi_u^t/\phi_p^t$ of the
density $\phi_u^t$ of all disks in the unpinned region to the
density $\phi_p^t$ of all disks in the pinned region for the
system in Fig.~\ref{fig:3} at different values of $F_M$.
Figure~\ref{fig:5}(b) shows the ratio $R_a=\phi_u^a/\phi_p^a$ of the density
of active disks in the unpinned and pinned regions, while in
Fig.~\ref{fig:5}(c) we plot the corresponding ratio $R_p=\phi_u^p/\phi_p^p$ for
the passive disks.
At initialization, we have $R_t=R_a=R_p=1.0$.
We first focus on the case of $F_{M} = 1.5$, which matches  
the system shown in Fig.~\ref{fig:2}.
There is an initial rapid drop in $R_t$ from $R_t=1.0$ to $R_t=0.4$
which is accompanied by
an even more rapid drop in the active particle ratio from $R_a=1.0$ to
$R_a\approx 0.1$. 
As the active disks enter the pinned region,
they drag along a portion of the passive disks,
as indicated
by the small drop in $R_p$
for $t < 0.05\times 10^7$ in Fig.~\ref{fig:5}(c). 
In Fig.~\ref{fig:6} we show a blow up
of Fig.~\ref{fig:5}
over the range $t<0.1\times 10^7$
in order to illustrate more clearly
the initial invasion of the active and passive disks
into the pinned region.
A portion of the passive disks are dragged into the pinned region
by the active disks, causing $R_p$ to decrease.
For $t > 0.05\times 10^7$, there is a crossover to the second stage
in which $R_a$ remains fixed but $R_p$ gradually increases due to
the shepherding process in which
the active disks eject the passive disks from the pinned region.
Figure~\ref{fig:5} shows that $R_p$ increases
from $R_p=0.9$ to $R_p \approx 2.5$.
At the same time, the density ratio for all disks $R_t$ increases
and
approaches $R_t=0.75$ at the longest times.
This suggests that for even longer times,
the system should gradually approach
a state with nearly uniform total density in which the active and
passive disks are phase separated in the pinned and unpinned regions.
In Fig.~\ref{fig:5} we find that the time needed to reach a
species separated state increases
with decreasing $F_{M}$.
The duration of the initial invasion of
the active particles into the pinned region also increases with
decreasing $F_M$, and
for $F_{M} < 0.9$ we resolve only
the initial invasion stage.

\begin{figure}
\includegraphics[width=\columnwidth]{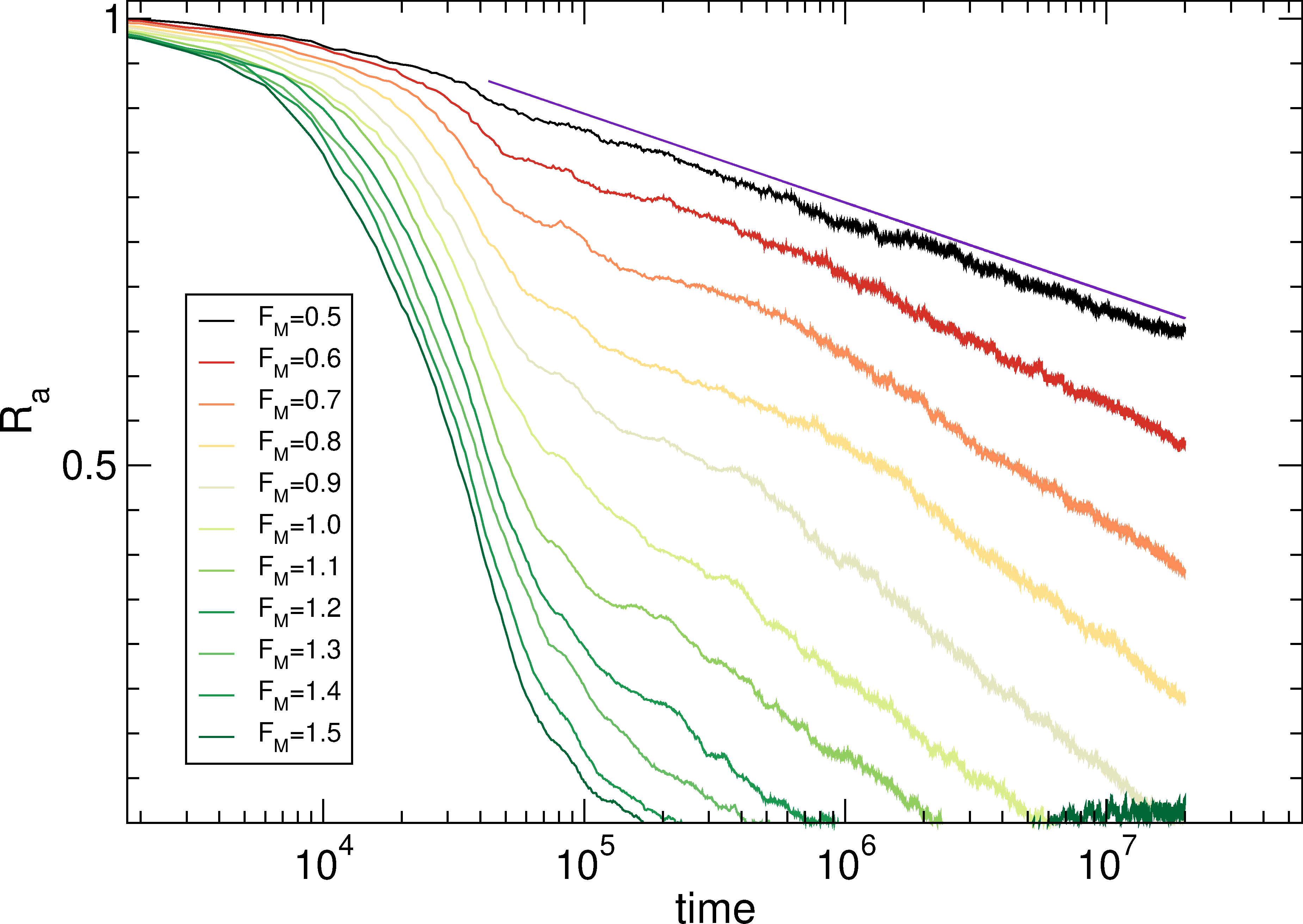}
\caption{Linear-log plot of the 
active disk density ratio $R_a$ versus time
for the system in Fig.~\ref{fig:5} with $N_p=9000$, $l_r=600$, and
varied $F_{M}$.
For lower motor forces the
curves decay approximately as
$R_a \propto A - B \log(t)$, as indicated by the upper solid line. 
}
\label{fig:7}
\end{figure}

In Fig.~\ref{fig:7} we plot $R_a$ versus time
on a linear-log scale to show more clearly
the slowing of the initial invasion stage with decreasing motor force.
The curves for $F_M<1.0$ can be described as having three parts.
In the first part
for $t < 2\times 10^5$,
there is an initial sharp decrease
in $R_a$ corresponding to the moving front of active disks.
This is followed by a slower relaxation of the form
$R_a \propto A - B\log(t)$, and
finally by
an eventual saturation close to a ratio $R_a=0.1$.
Logarithmic relaxation has been observed in 
vibrated granular matter \cite{Nowak98} where there
is a gradual compaction to a denser state.
In our system the equivalent of the
compaction dynamics
is the build up of the active disks in the pinned region.   

\begin{figure}
\includegraphics[width=\columnwidth]{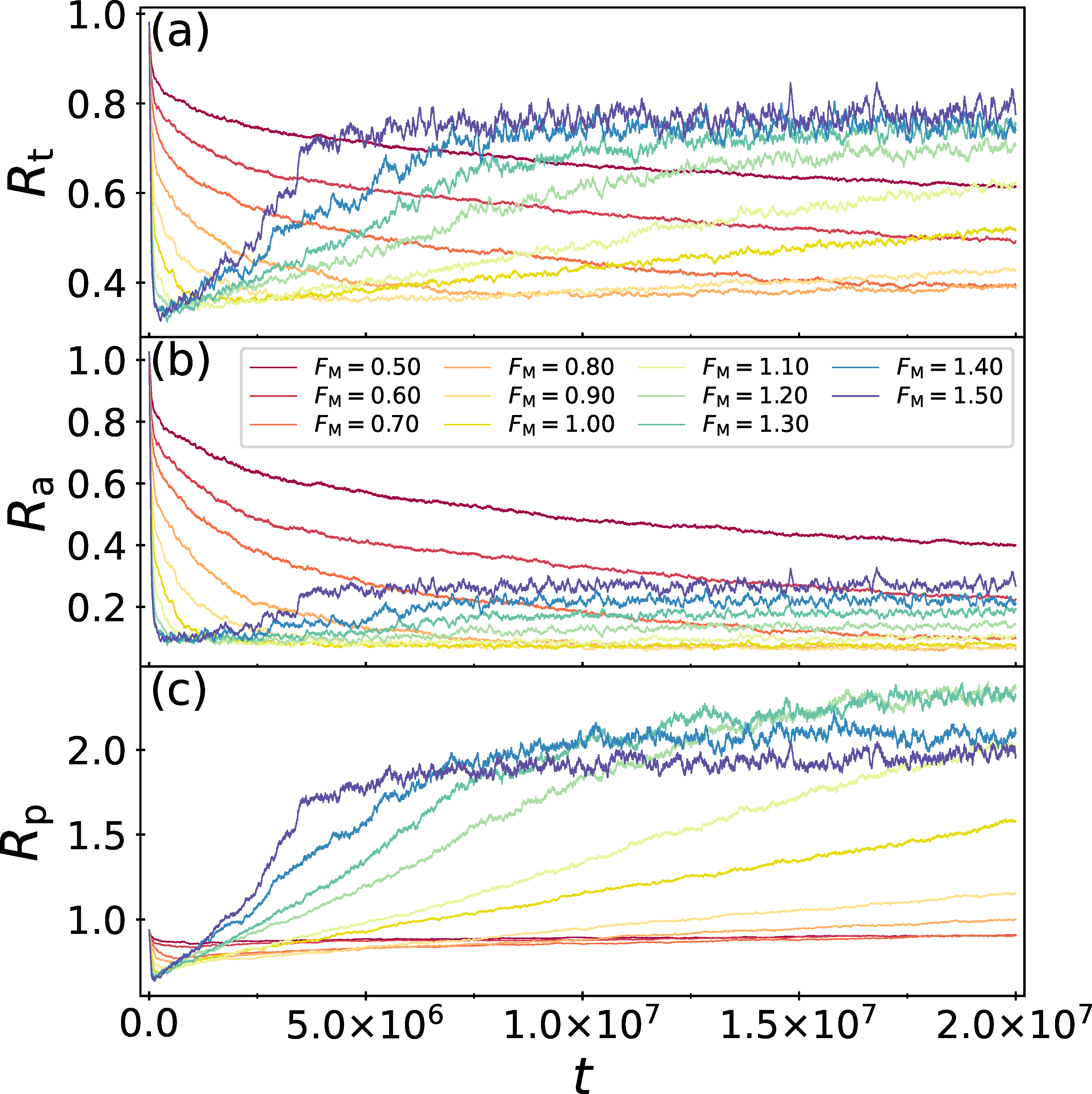}
\caption{
Density ratios for the unpinned to pinned regions for a system with
$N_p=6000$ and $l_r=600$
at varied $F_M$.
(a) The density ratio of all disks $R_t=\phi_u^t/\phi_p^t$.
(b) The active disk density ratio $R_a=\phi_u^a/\phi_p^a$.
(c) The passive disk density ratio $R_p=\phi_u^p/\phi_p^p$.
At high $F_M$,
the system saturates to a state with $R_t=0.75$.
}
\label{fig:8}
\end{figure}

In Fig.~\ref{fig:8}
we plot $R_t$, $R_a$, and $R_p$ versus time for a
system with a lower number of pinning sites $N_{p}=6000$ at different
values of $F_M$.
The overall behavior is similar to that found
for the $N_{p} = 9000$ sample.
When $F_{M} > 0.8$ we find
the two step
process of an
initial active disk invasion of the pinned region
and the later passive disk shepherding out of the pinned region. 
For $F_{M} > 1.2$, the
system reaches a steady
state with $R_t=0.75$,
indicating a higher overall density in the pinned region compared
to the unpinned region.

\begin{figure}
\includegraphics[width=\columnwidth]{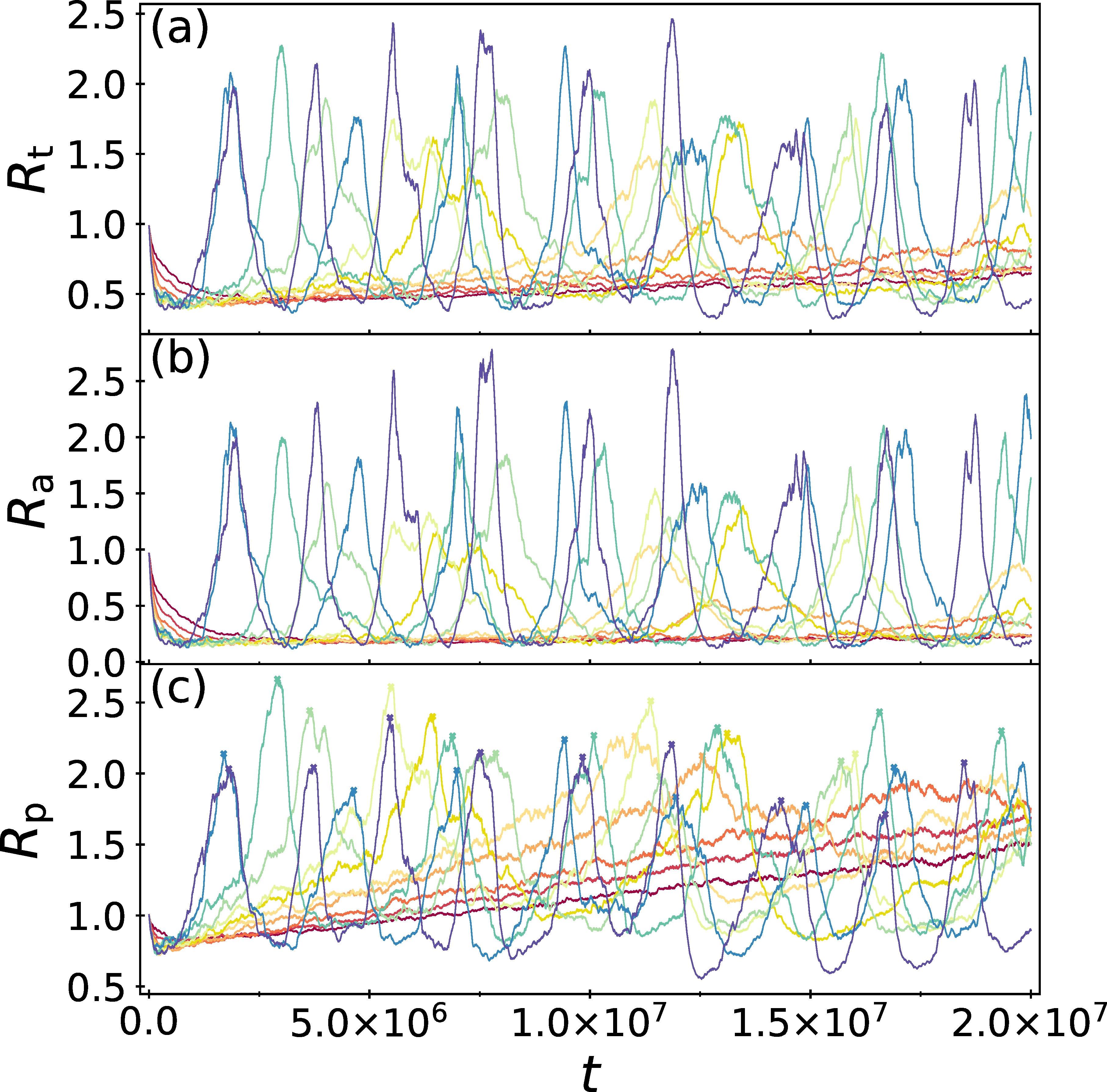}
\caption{
Density ratios
for the unpinned to pinned regions   versus time
for a system with
$N_p=2000$ and $l_r=600$
at $F_M=0.5$, 0.6, 0.7, 0.8, 0.9, 1.0, 1.1, 1.2, 1.3, 1.4, and 1.5, from
top left
to bottom left.
(a) The density ratio of all disks $R_t=\phi_u^t/\phi_p^t$.
(b) The active disk density ratio $R_a=\phi_u^a/\phi_p^a$.
(c) The passive disk density ratio $R_p=\phi_u^p/\phi_p^p$.
For $F_{M} >0.8$, we find strong density oscillations due to 
motility induced clustering,
as illustrated in Fig.~\ref{fig:10}(a).  
}
\label{fig:9}
\end{figure}

\begin{figure}
\includegraphics[width=\columnwidth]{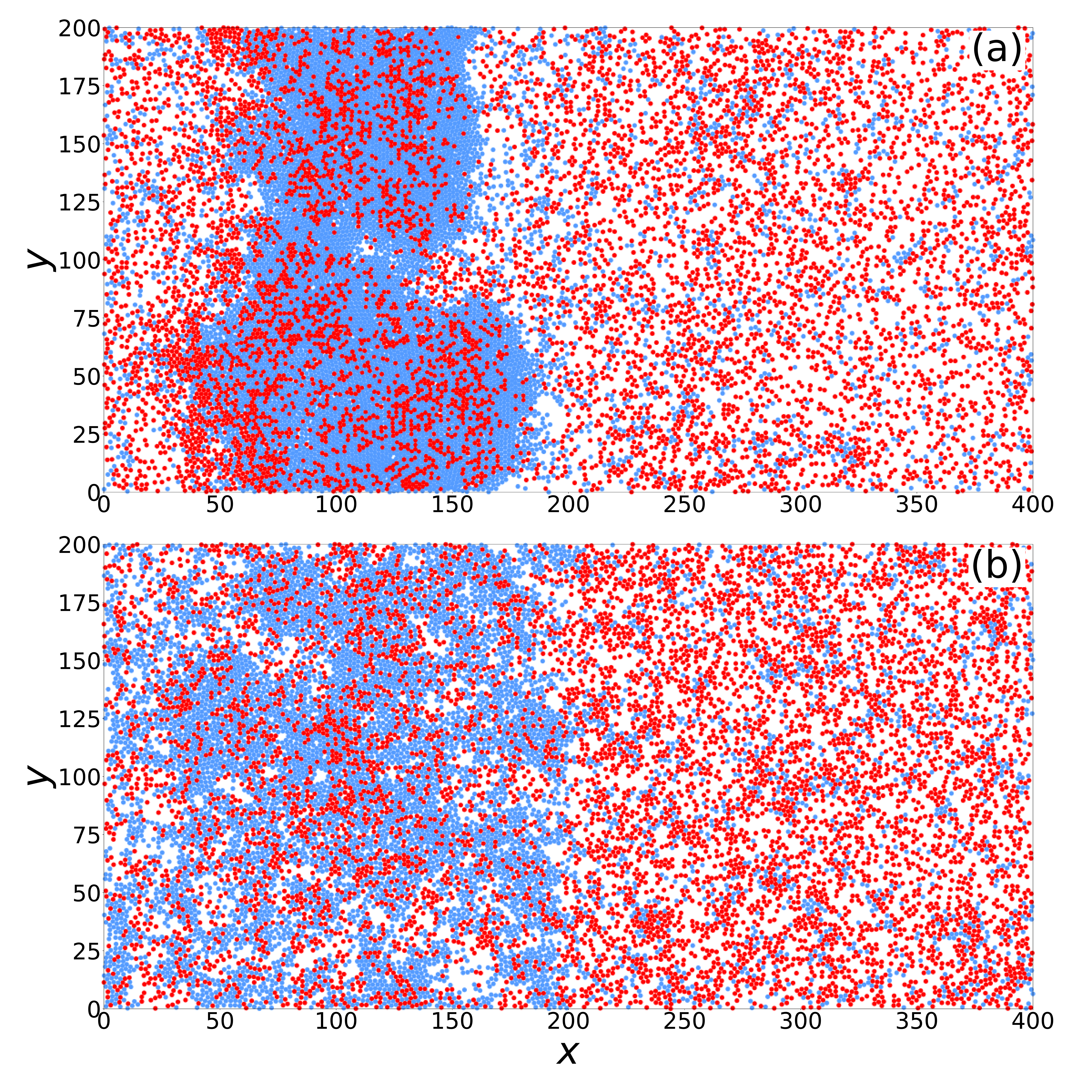}
\caption{
Images of the active disks (blue) and passive disks (red) for the system
in Fig.~\ref{fig:9} with $N_p=2000$ and $l_r=600$
at $t = 2\times 10^7$.
(a) For $F_{M} = 1.5$, the system forms a motility induced cluster state. 
(b) At $F_{M} = 0.5$, there is a liquid like state
with some species phase separation.  
}
\label{fig:10}
\end{figure}

In Fig.~\ref{fig:9} we plot $R_t$, $R_a$, and $R_p$ versus time for a system
with an even lower number of pinning sites $N_{p} = 2000$.
For $F_{M} < 0.8$
we find the usual two stage behavior
of a rapid advancement of active disks into the pinned region followed
by a slow
species phase separation at longer times.
For $F_{M} > 0.8$, the behavior changes and
there are strong oscillations in $R_t$, $R_a$, and $R_p$.
These oscillations result when a large scale
motility induced
clustering state
forms, as shown in
Fig.~\ref{fig:10}(a) at $t = 2\times 10^7$ for the system
in Fig.~\ref{fig:9}
at $F_{M} = 1.5$.
In the pinned region, there is
a background of pinned disks
coexisting with a much larger cluster
composed of both active and inactive disks.
The density ratio oscillations in Fig.~\ref{fig:9} are correlated
with each other, indicating that there is either no species phase separation
or at most only very weak separation.
In the snapshot, the large cluster
is located in the pinned region, but the cluster is
mobile and
spends an equal amount of time in the pinned and unpinned regions, producing
the density ratio oscillations and giving a time averaged density that is
close to uniform.
The motility induced 
cluster in Fig.~\ref{fig:10}(a) differs from the clustering
found in the pinned region at higher $N_{p}$,
since the latter clusters are mostly frozen in time and have little
to no mobility. 
In contrast, the cluster shown in Fig.~\ref{fig:10}(a) gradually drifts
through the entire sample over time.
In 
Fig.~\ref{fig:10}(b), we illustrate the disk configurations
for the system in Fig.~\ref{fig:9} for $F_M=0.5$ at
$t = 2\times 10^7$,
where the clustering is lost but there
is still some phase separation.
We have constructed
a series of plots similar to those shown in
Fig.~\ref{fig:9} and find that 
motility induced clustering
always occurs in samples with $N_{p} < 4500$
at the larger motor forces. 

\begin{figure}
\includegraphics[width=\columnwidth]{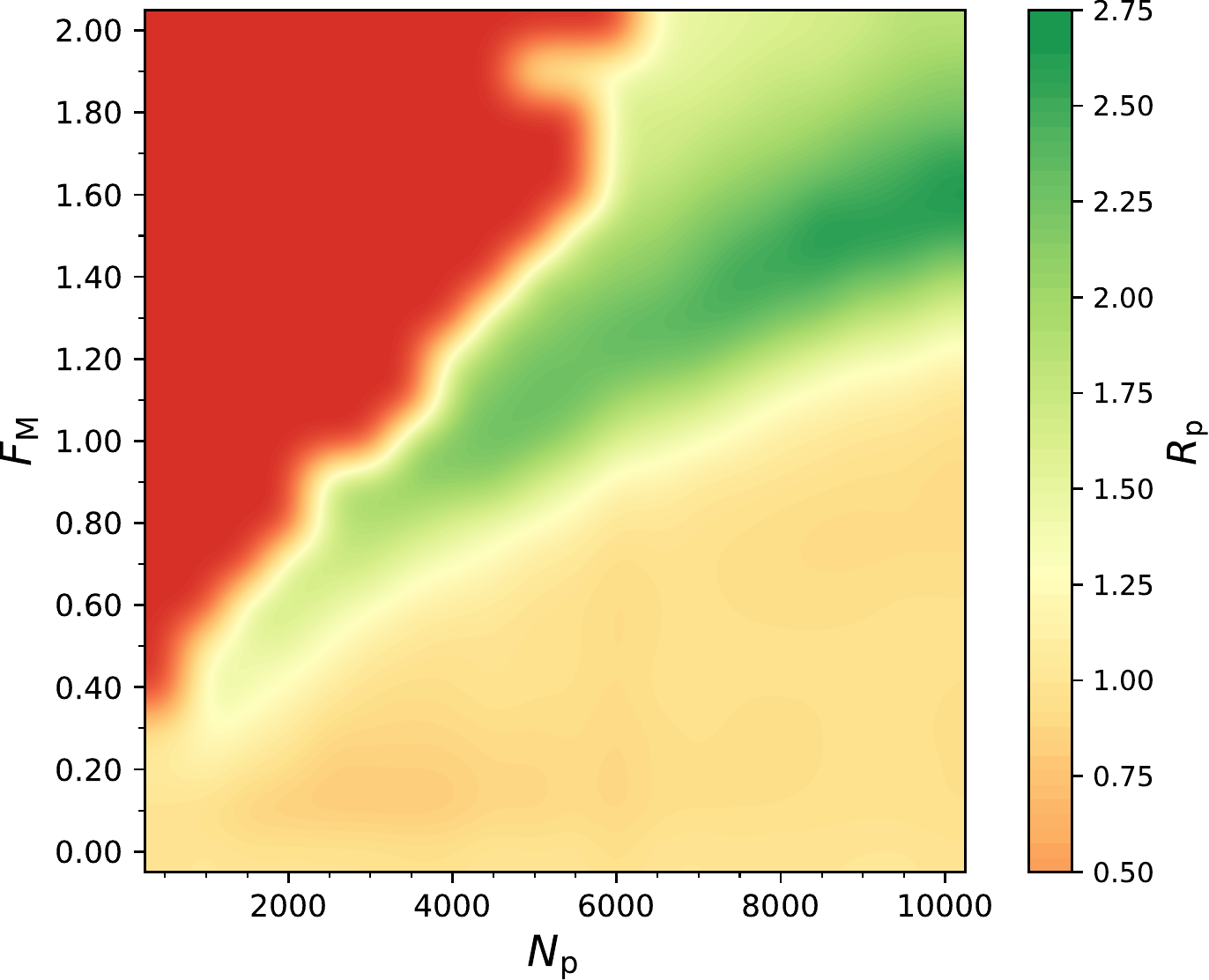}
\caption{
Dynamic phase diagram as a function of motor force $F_M$ versus number of
pinning sites $N_p$ obtained from the value of the passive disk density
ratio $R_p$  at a time $t=2\times 10^7$.
Green: 
strong shepherding with species separation.
Yellow: Weak or no shepherding.
Red: large density oscillations occur due to the formation of the
motility induced cluster state illustrated in Figs.~\ref{fig:9} and
\ref{fig:10}(a).
}
\label{fig:11}
\end{figure}

Based on the behavior of the density ratios and our determination of whether
density oscillations of the type shown in Fig.~\ref{fig:9} are present,
we can construct a dynamic phase
diagram of the different behaviors as
a function of $F_M$ versus $N_p$,
as shown in Fig.~\ref{fig:11} using the value of
the passive disk density ratio $R_p$ at a time
$t = 2\times 10^7$.
In the green region, we find strong shepherding
effects and the sample reaches a saturated state, while
in the yellow region, the shepherding is weak or absent.
The motility induced cluster state of the type illustrated
in Fig.~\ref{fig:10}(a) appears in the red region.
We find that shepherding can only
occur for a sufficiently large number of pinning sites and a sufficiently
high motor force.
The motility induced cluster state
grows in extent as $F_M$ increases.
When the number of pins is large
but $F_{M}$ is small,
the pinned portion of the sample is in a
disordered or glassy state.
In this study we fix $F_p=2.5$ and 
consider only the regime $F_M<F_p$,
but we expect that when
$F_{M} > F_{p}$, the motility induced cluster state
will
dominate the phase diagram. 

\begin{figure}
\includegraphics[width=\columnwidth]{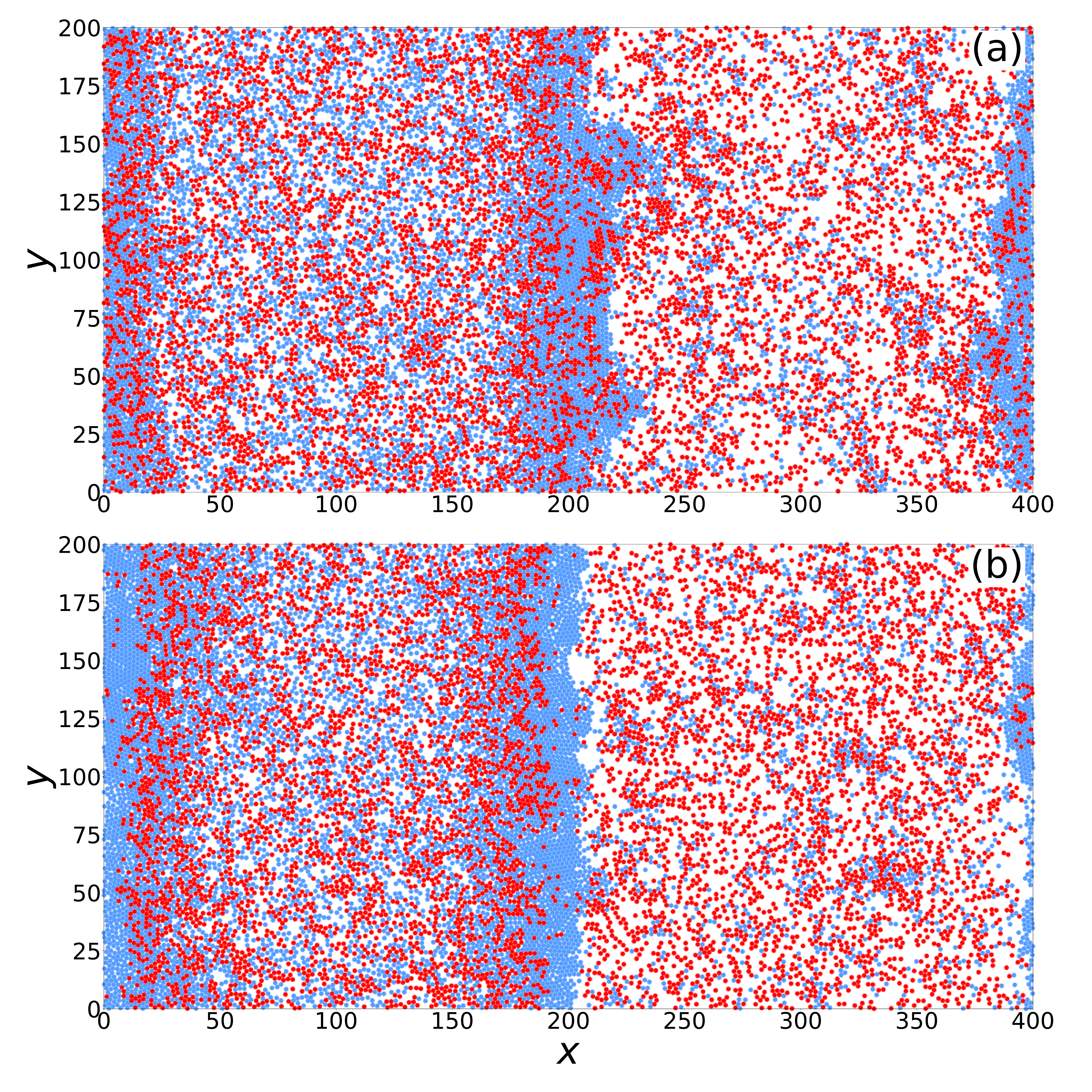}
\caption{
Images of the active disks (blue) and passive disks (red) for a system
with $N_p=8500$, $l_r=600$, and $F_M=0.7$,
near the transition between the shepherding
and non-shepherding states in Fig.~\ref{fig:11}.
(a) 
At $t = 5\times 10^5$,
a dense interface of active disks has formed at the edge of the
pinned region.
(b) The image at a much later time of
$t=2\times 10^7$ shows that this interface is effectively pinned.
}
\label{fig:12}
\end{figure}

At intermediate motor forces and larger $N_p$,
near the phase boundary between shepherding (green) and
absence of shepherding (yellow) in Fig.~\ref{fig:11},
a frozen invasion front can appear,
as illustrated in Fig.~\ref{fig:12} for
a sample with $N_p = 8500$ and $F_{M} = 0.7$.
Here the initial invasion of the active front occurs slowly enough that
as the front starts to penetrate the pinned region, the front itself
becomes pinned,
forming a crystallized state near the edge
of the pinned region.
This pinned dense front
makes it difficult for the passive disks to escape
into the unpinned region.
Figure~\ref{fig:12}(b) shows that at
a much later time of $t=2\times 10^7$,
the dense front has become better defined
and has failed to penetrate further into the pinned region.
The effective drive experienced by the dense front
is proportional to $F_{M}$.
In other systems where an interface moves over quenched disorder,
it is known that there is
is a minimum critical force $F_{c}$ which must be applied in order for the
interface to depin \cite{Reichhardt17},
so the crossover from the shepherding regime to the
non-shepherding glassy state can be viewed as occurring at the point
where $F_{M}$ drops below the effective $F_{c}$ for interface depinning.

There have been several studies examining bidisperse
mixtures of passive and active particles which have shown that
both motility induced phase separation and species separation
can occur
\cite{Stenhammar15,Ai18,Kolb20,Rodriguez20}.
It has also been demonstrated that
active particles can move passive particles
and organize them into particular 
states even when there are only a small number of active particles
present, which is known as active doping
\cite{Ni14,Kummel15,Ramananarivo19,Omar19}.
This is similar to the shepherding
phenomenon we observe. 
Our results show that quenched disorder could be used to
achieve both density and species phase separation. For example,
it should be possible to
create a patterned state with pinning
where the active particles could accumulate and
gradually move the inactivate particles
in order to create
a tailored self-assembled arrangement of active and passive particles.

\begin{figure}
\includegraphics[width=\columnwidth]{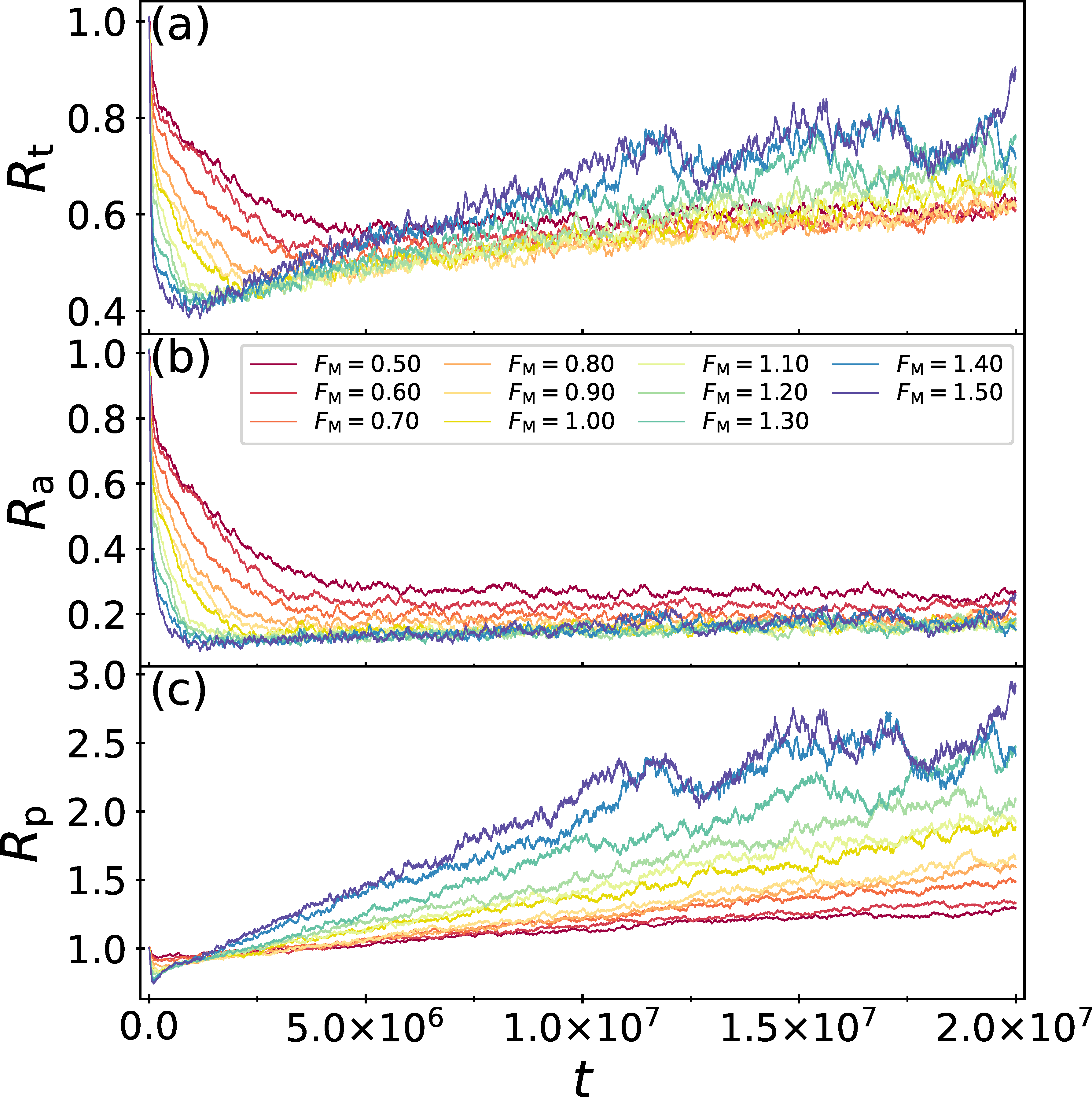}
\caption{Density ratios for the unpinned to pinned regions
versus time for a system
with $N_{\rm obs}=1500$ repulsive obstacles
and $l_r=600$.
(a) The density ratio of all disks $R_t=\phi_u^t/\phi_p^t$.
(b) The active disk density ratio $R_a=\phi_u^a/\phi_p^a$.
(c) The passive disk density ratio $R_p=\phi_u^p/\phi_p^p$.
Here we find that
the active disks move rapidly into the obstacle-filled region and then
gradually shepherd the passive particles into the obstacle-free region.
 }
\label{fig:13}
\end{figure}

\section{Obstacle Arrays}
We have also considered samples in which the attractive pinning sites are
replaced by repulsive obstacles in the form of immobile disks.
In Fig.~\ref{fig:13}(a,b,c) we plot $R_t$, $R_a$, and $R_p$ as a function
of time for a system with $N_{\rm obs}=1500$ obstacles and
$F_{M} = 1.5$.
The active disks rapidly move into the obstacle region,
followed by the 
long time shepherding of the inactive disks into the unpinned region,
similar to what we find for samples with attractive pinning sites at
much higher pinning densities of $N_p>5000$.

\begin{figure}
\includegraphics[width=\columnwidth]{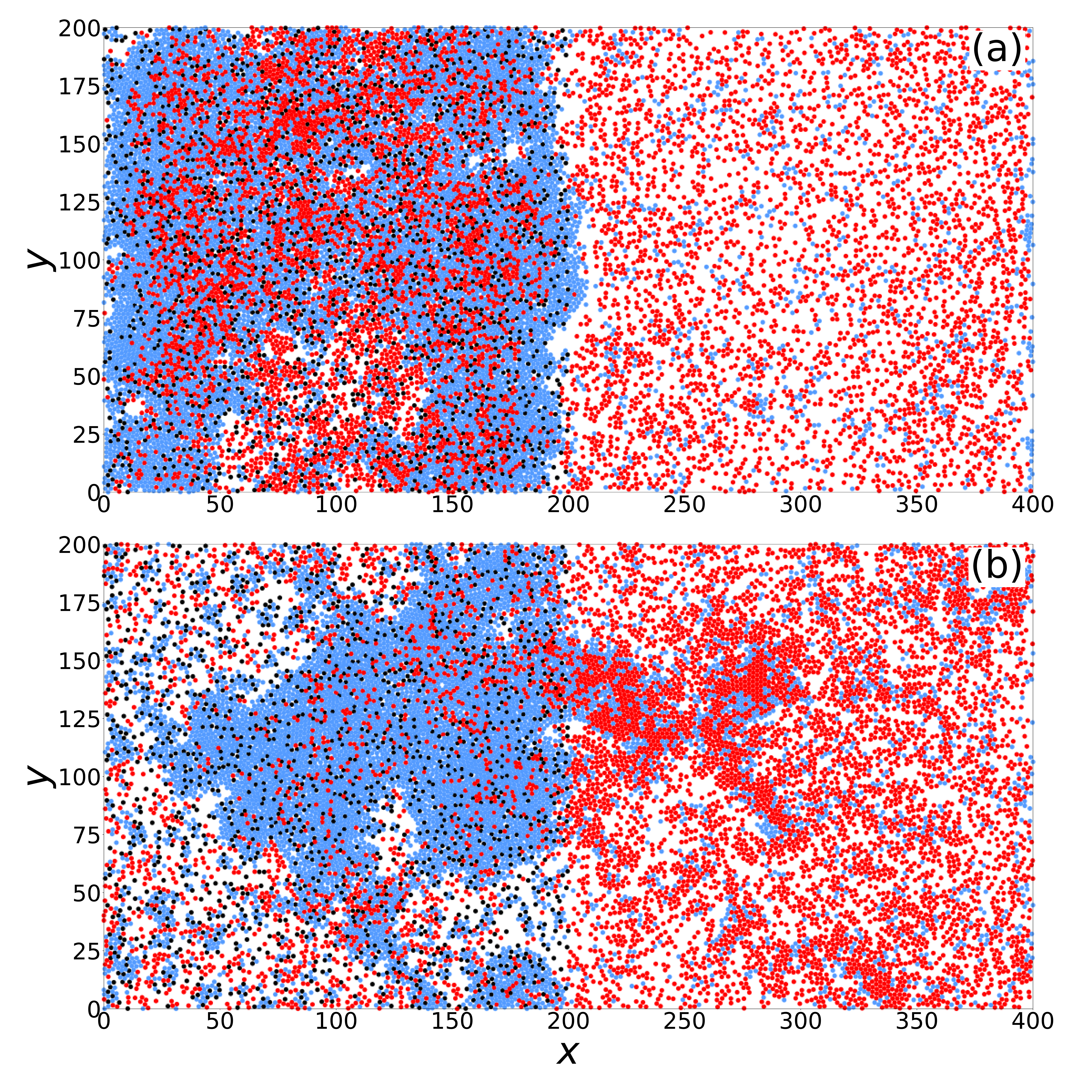}
\caption{
Images of the active disks (blue), passive disks (red), and obstacle locations
(black) for the system in Fig.~\ref{fig:13} with repulsive obstacles instead
of attractive pinning sites at $N_{\rm obs}=1500$, $l_r=600$,
and $F_M=1.5$.
(a) At $t = 0.05\times 10^7$, there is an initial invasion
into the region with obstacles.
(b) At $t = 2\times 10^7$, there is a stronger species separation.
 }
\label{fig:14}
\end{figure}

In Fig.~\ref{fig:14}(a) we show the positions of the active and passive
disks along with the obstacle locations
for the system in Fig.~\ref{fig:13} at $t = 0.05\times 10^7$, where the
active and passive disks begin to cluster in the obstacle-filled region.
In Fig.~\ref{fig:14}(b), the same system at a later time
of $t = 2.0\times 10^7$ has
a stronger species separation.
In the obstacle-filled region,
the active disk density is higher
than in samples containing pinning sites. 
Previous work for active matter on random pinning arrays
showed
that there can be a wetting transition
in which mobility induced clusters break apart
as the active particles spread out and attempt to
occupy as many pinning sites as possible
\cite{Sandor17b}.
On the other hand, it has also been shown
that
repulsive obstacles can act as
nucleation sites promoting the clustering
of active particles \cite{Reichhardt14a}.
The clusters which are induced by the presence of the obstacles
remain pinned at the locations of the obstacles and are not mobile.
In our system with obstacles,
we do not observe
drifting or floating clusters of the type illustrated
in Fig.~\ref{fig:10}(a) for
pinning site systems with
low $N_p$
and high $F_{M}$,
since even a small number of obstacles can pin
down any cluster that forms,
while
the cluster can effectively float above the pinning sites if $F_M$ is
sufficiently large.
In the obstacle-free region
we observe some
weak clustering due to the shepherding of the passive particles by a
small number of active particles,
as shown in Fig.~\ref{fig:14}(b).
We find behavior similar to that illustrated in Figs.~\ref{fig:13} and
\ref{fig:14} for other obstacle densities.

\section{Monodisperse active disks}

\begin{figure}
\includegraphics[width=\columnwidth]{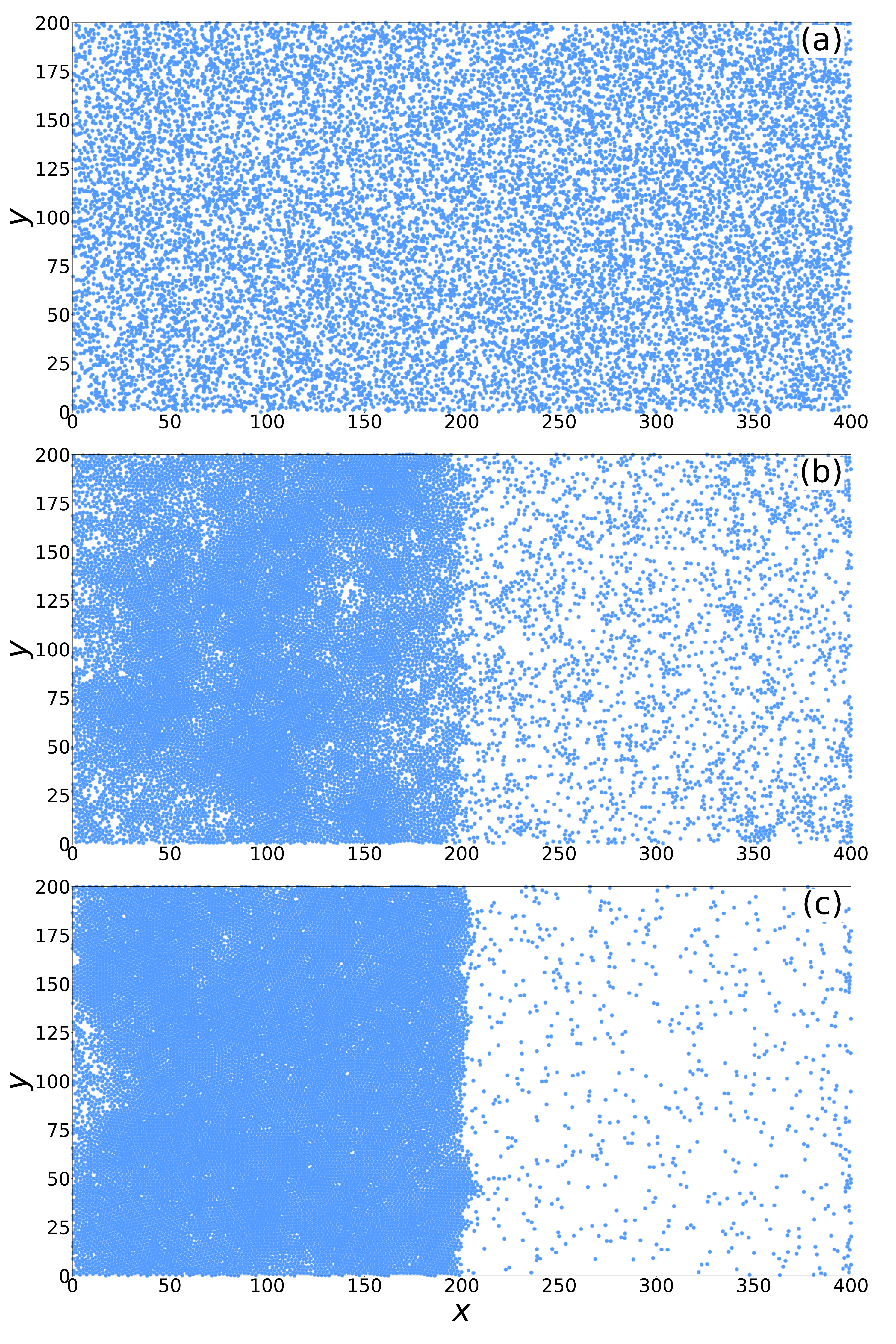}
\caption{
Images of the active disks (blue) for a system with $N_p=9000$ pinning
sites and no passive disks at $l_r=600$ and $F_M=1.5$.
(a) The $t=0$ initial uniform configuration.
(b) At $t=0.02\times 10^7$,
clusters form on both sides of the pinned region.
(c) The fully density phase separated state at
$t=2.0\times 10^7$.
}
\label{fig:15}
\end{figure}

We have also considered monodisperse systems in which all of the disks
are active and there are no passive disks.
Here,
we find only two generic phases. The first is
the accumulation of active disks
in the pinned region, forming a large density gradient,
and the second is the formation of
a large scale drifting cluster that floats over the
random pinning. 
In Fig.~\ref{fig:15}(a) we show the
initial uniform configuration for a monodisperse system containing
only active disks with $N_{p} = 9000$ pinning sites and $F_{M} = 1.5$.
At $t = 0.01\times 10^7$
in Fig.~\ref{fig:15}(b),
mobile clusters have formed in the unpinned region
while 
pinned clusters appear
in the pinned region.
In Fig.~\ref{fig:15}(c)
at $t = 2\times 10^7$,
most of the disks are in the pinned region with
a low density gas of active disks
present in the unpinned region.
For smaller $F_{M}$, the same behavior occurs
but the time required for the active disks to
penetrate the pinned region fully increases.
  
If we place monodisperse active disks in a sample containing obstacles
instead of pinning sites, we find behavior similar to that shown
in Fig.~\ref{fig:14},
but
the mobile cluster phase is absent since the
clusters become pinned in the obstacle-filled region.
In general, one might expect the active disks
to accumulate in the unpinned region where they have more space to move;
however, this is not what occurs due to a combination of effects.
The first is the reduction
of the mobility through direct interactions between the active disks and
the pinning sites or obstacles,
and the second is the motility induced
clustering effect 
which slows down the disks and is enhanced by the presence of pinning sites
or obstacles.
In a system where one region has a low diffusion coefficient and another
region has a high diffusion coefficient,
particles in the high diffusion region can
rapidly 
explore space and reach the low diffusion region, whereas particles in
the low diffusion region have reduced mobility and cannot reach the high
diffusion region easily. The resulting flux imbalance causes the low diffusion
region to accumulate a higher concentration of particles.
In the case of active disks,
the motility induced clustering state
also produces a lower effective diffusion coefficient compared to
freely moving non-interacting active disks.
If a landscape structure could be identified
which breaks apart the motility induced clusters
only in the disordered region
but not in the disorder-free region,  it would
be possible to achieve a higher effective diffusion coefficient in the
disordered region, causing an accumulation of particles in the unpinned
region, as opposed to what we observe, which is
the opposite effect where the
active disks accumulate in the pinned or obstacle-filled region.

\section{Summary}
We have examined a bidisperse mixture of active and passive disks
interacting with inhomogeneous disorder in which
half of the sample contains randomly distributed
pinning sites and the other
half is free of disorder.
We consider active disks obeying run-and-tumble dynamics,
and vary the magnitude of the motor force and
the density of pinning sites.
For dense pinning and high motor forces,
we observe a two step process in which active and passive particles
accumulate in the pinned region, producing a large density gradient,
followed by
a slower shepherding process in which
the passive disks are pushed into the unpinned region, producing
a state
that is more uniform in density 
but is phase separated.
For lower motor forces, the initial invasion process becomes slower,
and if the motor force is below a critical value, 
the invasion front entering the pinned region becomes pinned. 
For larger motor forces and lower pinning density,
we find large scale drifting clusters containing a mixture of active and
passive disks which
effectively float over
the pinning sites,
leading to large oscillations in the density on either side of the sample
but giving a uniform time averaged density across the entire sample.
If we replace the attractive pinning sites by repulsive obstacles,
the drifting cluster
phase is lost and
we observe an expanded shepherding region
with the formation of much more compact clusters in the disordered 
portion of the sample.
For a monodisperse 
system containing only active disks and no passive disks,
the disks accumulate in the pinned region
for higher pinning density and
form a drifting motility induced cluster state
at low pinning densities.
Our results indicate that active particles
could be used to move passive particles though complex landscapes or
to control the invasion of active fluids into 
disordered media.    

\begin{acknowledgments}
This work was supported by the US Department of Energy through
the Los Alamos National Laboratory.  Los Alamos National Laboratory is
operated by Triad National Security, LLC, for the National Nuclear Security
Administration of the U. S. Department of Energy (Contract No. 892333218NCA000001).
PF and AL were supported by a grant of the Romanian Ministry of Education
and Research, CNCS - UEFISCDI, project number
PN-III-P4-ID-PCE-2020-1301, within PNCDI III.
\end{acknowledgments}

\bibliography{mybib}

\begin{thebibliography}{68}%
\makeatletter
\providecommand \@ifxundefined [1]{%
 \@ifx{#1\undefined}
}%
\providecommand \@ifnum [1]{%
 \ifnum #1\expandafter \@firstoftwo
 \else \expandafter \@secondoftwo
 \fi
}%
\providecommand \@ifx [1]{%
 \ifx #1\expandafter \@firstoftwo
 \else \expandafter \@secondoftwo
 \fi
}%
\providecommand \natexlab [1]{#1}%
\providecommand \enquote  [1]{``#1''}%
\providecommand \bibnamefont  [1]{#1}%
\providecommand \bibfnamefont [1]{#1}%
\providecommand \citenamefont [1]{#1}%
\providecommand \href@noop [0]{\@secondoftwo}%
\providecommand \href [0]{\begingroup \@sanitize@url \@href}%
\providecommand \@href[1]{\@@startlink{#1}\@@href}%
\providecommand \@@href[1]{\endgroup#1\@@endlink}%
\providecommand \@sanitize@url [0]{\catcode `\\12\catcode `\$12\catcode
  `\&12\catcode `\#12\catcode `\^12\catcode `\_12\catcode `\%12\relax}%
\providecommand \@@startlink[1]{}%
\providecommand \@@endlink[0]{}%
\providecommand \url  [0]{\begingroup\@sanitize@url \@url }%
\providecommand \@url [1]{\endgroup\@href {#1}{\urlprefix }}%
\providecommand \urlprefix  [0]{URL }%
\providecommand \Eprint [0]{\href }%
\providecommand \doibase [0]{http://dx.doi.org/}%
\providecommand \selectlanguage [0]{\@gobble}%
\providecommand \bibinfo  [0]{\@secondoftwo}%
\providecommand \bibfield  [0]{\@secondoftwo}%
\providecommand \translation [1]{[#1]}%
\providecommand \BibitemOpen [0]{}%
\providecommand \bibitemStop [0]{}%
\providecommand \bibitemNoStop [0]{.\EOS\space}%
\providecommand \EOS [0]{\spacefactor3000\relax}%
\providecommand \BibitemShut  [1]{\csname bibitem#1\endcsname}%
\let\auto@bib@innerbib\@empty
\bibitem [{\citenamefont {Marchetti}\ \emph {et~al.}(2013)\citenamefont
  {Marchetti}, \citenamefont {Joanny}, \citenamefont {Ramaswamy}, \citenamefont
  {Liverpool}, \citenamefont {Prost}, \citenamefont {Rao},\ and\ \citenamefont
  {Simha}}]{Marchetti13}%
  \BibitemOpen
  \bibfield  {author} {\bibinfo {author} {\bibfnamefont {M.~C.}\ \bibnamefont
  {Marchetti}}, \bibinfo {author} {\bibfnamefont {J.~F.}\ \bibnamefont
  {Joanny}}, \bibinfo {author} {\bibfnamefont {S.}~\bibnamefont {Ramaswamy}},
  \bibinfo {author} {\bibfnamefont {T.~B.}\ \bibnamefont {Liverpool}}, \bibinfo
  {author} {\bibfnamefont {J.}~\bibnamefont {Prost}}, \bibinfo {author}
  {\bibfnamefont {M.}~\bibnamefont {Rao}}, \ and\ \bibinfo {author}
  {\bibfnamefont {R.~A.}\ \bibnamefont {Simha}},\ }\bibfield  {title} {\enquote
  {\bibinfo {title} {Hydrodynamics of soft active matter},}\ }\href {\doibase
  10.1103/RevModPhys.85.1143} {\bibfield  {journal} {\bibinfo  {journal} {Rev.
  Mod. Phys.}\ }\textbf {\bibinfo {volume} {85}},\ \bibinfo {pages}
  {1143--1189} (\bibinfo {year} {2013})}\BibitemShut {NoStop}%
\bibitem [{\citenamefont {Bechinger}\ \emph {et~al.}(2016)\citenamefont
  {Bechinger}, \citenamefont {Di~Leonardo}, \citenamefont {L\"owen},
  \citenamefont {Reichhardt}, \citenamefont {Volpe},\ and\ \citenamefont
  {Volpe}}]{Bechinger16}%
  \BibitemOpen
  \bibfield  {author} {\bibinfo {author} {\bibfnamefont {C.}~\bibnamefont
  {Bechinger}}, \bibinfo {author} {\bibfnamefont {R.}~\bibnamefont
  {Di~Leonardo}}, \bibinfo {author} {\bibfnamefont {H.}~\bibnamefont
  {L\"owen}}, \bibinfo {author} {\bibfnamefont {C.}~\bibnamefont {Reichhardt}},
  \bibinfo {author} {\bibfnamefont {G.}~\bibnamefont {Volpe}}, \ and\ \bibinfo
  {author} {\bibfnamefont {G.}~\bibnamefont {Volpe}},\ }\bibfield  {title}
  {\enquote {\bibinfo {title} {Active particles in complex and crowded
  environments},}\ }\href {\doibase 10.1103/RevModPhys.88.045006} {\bibfield
  {journal} {\bibinfo  {journal} {Rev. Mod. Phys.}\ }\textbf {\bibinfo {volume}
  {88}},\ \bibinfo {pages} {045006} (\bibinfo {year} {2016})}\BibitemShut
  {NoStop}%
\bibitem [{\citenamefont {Gompper}\ \emph {et~al.}(2020)\citenamefont
  {Gompper}, \citenamefont {Winkler}, \citenamefont {Speck}, \citenamefont
  {Solon}, \citenamefont {Nardini}, \citenamefont {Peruani}, \citenamefont
  {L{\" o}wen}, \citenamefont {Golestanian}, \citenamefont {Benjamin~Kaupp},
  \citenamefont {Alvarez}, \citenamefont {Ki{\o}rboe}, \citenamefont {Lauga},
  \citenamefont {Poon}, \citenamefont {DeSimone}, \citenamefont {Mui{\~
  n}os-Landin}, \citenamefont {Fischer}, \citenamefont {S{\" o}ker},
  \citenamefont {Cichos}, \citenamefont {Kapral}, \citenamefont {Gaspard},
  \citenamefont {Ripoll}, \citenamefont {Sagues}, \citenamefont
  {Doostmohammadi}, \citenamefont {Yeomans}, \citenamefont {Aranson},
  \citenamefont {Bechinger}, \citenamefont {Stark}, \citenamefont {Hemelrijk},
  \citenamefont {Nedelec}, \citenamefont {Sarkar}, \citenamefont {Aryaksama},
  \citenamefont {Lacroix}, \citenamefont {Duclos}, \citenamefont {Yashunsky},
  \citenamefont {Silberzan}, \citenamefont {Arroyo},\ and\ \citenamefont
  {Kale}}]{Gompper20}%
  \BibitemOpen
  \bibfield  {author} {\bibinfo {author} {\bibfnamefont {G.}~\bibnamefont
  {Gompper}}, \bibinfo {author} {\bibfnamefont {R.~G.}\ \bibnamefont
  {Winkler}}, \bibinfo {author} {\bibfnamefont {T.}~\bibnamefont {Speck}},
  \bibinfo {author} {\bibfnamefont {A.}~\bibnamefont {Solon}}, \bibinfo
  {author} {\bibfnamefont {C.}~\bibnamefont {Nardini}}, \bibinfo {author}
  {\bibfnamefont {F.}~\bibnamefont {Peruani}}, \bibinfo {author} {\bibfnamefont
  {H.}~\bibnamefont {L{\" o}wen}}, \bibinfo {author} {\bibfnamefont
  {R.}~\bibnamefont {Golestanian}}, \bibinfo {author} {\bibfnamefont
  {U.}~\bibnamefont {Benjamin~Kaupp}}, \bibinfo {author} {\bibfnamefont
  {L.}~\bibnamefont {Alvarez}}, \bibinfo {author} {\bibfnamefont
  {T.}~\bibnamefont {Ki{\o}rboe}}, \bibinfo {author} {\bibfnamefont
  {E.}~\bibnamefont {Lauga}}, \bibinfo {author} {\bibfnamefont {W.~C.~K.}\
  \bibnamefont {Poon}}, \bibinfo {author} {\bibfnamefont {A.}~\bibnamefont
  {DeSimone}}, \bibinfo {author} {\bibfnamefont {S.}~\bibnamefont {Mui{\~
  n}os-Landin}}, \bibinfo {author} {\bibfnamefont {A.}~\bibnamefont {Fischer}},
  \bibinfo {author} {\bibfnamefont {N.~A.}\ \bibnamefont {S{\" o}ker}},
  \bibinfo {author} {\bibfnamefont {F.}~\bibnamefont {Cichos}}, \bibinfo
  {author} {\bibfnamefont {R.}~\bibnamefont {Kapral}}, \bibinfo {author}
  {\bibfnamefont {P.}~\bibnamefont {Gaspard}}, \bibinfo {author} {\bibfnamefont
  {M.}~\bibnamefont {Ripoll}}, \bibinfo {author} {\bibfnamefont
  {F.}~\bibnamefont {Sagues}}, \bibinfo {author} {\bibfnamefont
  {A.}~\bibnamefont {Doostmohammadi}}, \bibinfo {author} {\bibfnamefont
  {Y.~M.}\ \bibnamefont {Yeomans}}, \bibinfo {author} {\bibfnamefont {I.~S.}\
  \bibnamefont {Aranson}}, \bibinfo {author} {\bibfnamefont {C.}~\bibnamefont
  {Bechinger}}, \bibinfo {author} {\bibfnamefont {H.}~\bibnamefont {Stark}},
  \bibinfo {author} {\bibfnamefont {C.~K.}\ \bibnamefont {Hemelrijk}}, \bibinfo
  {author} {\bibfnamefont {F.~J.}\ \bibnamefont {Nedelec}}, \bibinfo {author}
  {\bibfnamefont {T.}~\bibnamefont {Sarkar}}, \bibinfo {author} {\bibfnamefont
  {T.}~\bibnamefont {Aryaksama}}, \bibinfo {author} {\bibfnamefont
  {M.}~\bibnamefont {Lacroix}}, \bibinfo {author} {\bibfnamefont
  {G.}~\bibnamefont {Duclos}}, \bibinfo {author} {\bibfnamefont
  {V.}~\bibnamefont {Yashunsky}}, \bibinfo {author} {\bibfnamefont
  {P.}~\bibnamefont {Silberzan}}, \bibinfo {author} {\bibfnamefont
  {M.}~\bibnamefont {Arroyo}}, \ and\ \bibinfo {author} {\bibfnamefont
  {S.}~\bibnamefont {Kale}},\ }\bibfield  {title} {\enquote {\bibinfo {title}
  {The 2020 motile active matter roadmap},}\ }\href {\doibase
  10.1088/1361-648X/ab6348} {\bibfield  {journal} {\bibinfo  {journal} {J.
  Phys.: Condens. Matter}\ }\textbf {\bibinfo {volume} {32}},\ \bibinfo {pages}
  {193001} (\bibinfo {year} {2020})}\BibitemShut {NoStop}%
\bibitem [{\citenamefont {Fily}\ and\ \citenamefont
  {Marchetti}(2012)}]{Fily12}%
  \BibitemOpen
  \bibfield  {author} {\bibinfo {author} {\bibfnamefont {Y.}~\bibnamefont
  {Fily}}\ and\ \bibinfo {author} {\bibfnamefont {M.~C.}\ \bibnamefont
  {Marchetti}},\ }\bibfield  {title} {\enquote {\bibinfo {title} {Athermal
  phase separation of self-propelled particles with no alignment},}\ }\href
  {\doibase 10.1103/PhysRevLett.108.235702} {\bibfield  {journal} {\bibinfo
  {journal} {Phys. Rev. Lett.}\ }\textbf {\bibinfo {volume} {108}},\ \bibinfo
  {pages} {235702} (\bibinfo {year} {2012})}\BibitemShut {NoStop}%
\bibitem [{\citenamefont {Redner}\ \emph {et~al.}(2013)\citenamefont {Redner},
  \citenamefont {Hagan},\ and\ \citenamefont {Baskaran}}]{Redner13}%
  \BibitemOpen
  \bibfield  {author} {\bibinfo {author} {\bibfnamefont {G.~S.}\ \bibnamefont
  {Redner}}, \bibinfo {author} {\bibfnamefont {M.~F.}\ \bibnamefont {Hagan}}, \
  and\ \bibinfo {author} {\bibfnamefont {A.}~\bibnamefont {Baskaran}},\
  }\bibfield  {title} {\enquote {\bibinfo {title} {Structure and dynamics of a
  phase-separating active colloidal fluid},}\ }\href {\doibase
  10.1103/PhysRevLett.110.055701} {\bibfield  {journal} {\bibinfo  {journal}
  {Phys. Rev. Lett.}\ }\textbf {\bibinfo {volume} {110}},\ \bibinfo {pages}
  {055701} (\bibinfo {year} {2013})}\BibitemShut {NoStop}%
\bibitem [{\citenamefont {Palacci}\ \emph {et~al.}(2013)\citenamefont
  {Palacci}, \citenamefont {Sacanna}, \citenamefont {Steinberg}, \citenamefont
  {Pine},\ and\ \citenamefont {Chaikin}}]{Palacci13}%
  \BibitemOpen
  \bibfield  {author} {\bibinfo {author} {\bibfnamefont {J.}~\bibnamefont
  {Palacci}}, \bibinfo {author} {\bibfnamefont {S.}~\bibnamefont {Sacanna}},
  \bibinfo {author} {\bibfnamefont {A.~P.}\ \bibnamefont {Steinberg}}, \bibinfo
  {author} {\bibfnamefont {D.~J.}\ \bibnamefont {Pine}}, \ and\ \bibinfo
  {author} {\bibfnamefont {P.~M.}\ \bibnamefont {Chaikin}},\ }\bibfield
  {title} {\enquote {\bibinfo {title} {Living crystals of light-activated
  colloidal surfers},}\ }\href {\doibase 10.1126/science.1230020} {\bibfield
  {journal} {\bibinfo  {journal} {Science}\ }\textbf {\bibinfo {volume}
  {339}},\ \bibinfo {pages} {936--940} (\bibinfo {year} {2013})}\BibitemShut
  {NoStop}%
\bibitem [{\citenamefont {Buttinoni}\ \emph {et~al.}(2013)\citenamefont
  {Buttinoni}, \citenamefont {Bialk\'e}, \citenamefont {K\"ummel},
  \citenamefont {L\"owen}, \citenamefont {Bechinger},\ and\ \citenamefont
  {Speck}}]{Buttinoni13}%
  \BibitemOpen
  \bibfield  {author} {\bibinfo {author} {\bibfnamefont {I.}~\bibnamefont
  {Buttinoni}}, \bibinfo {author} {\bibfnamefont {J.}~\bibnamefont {Bialk\'e}},
  \bibinfo {author} {\bibfnamefont {F.}~\bibnamefont {K\"ummel}}, \bibinfo
  {author} {\bibfnamefont {H.}~\bibnamefont {L\"owen}}, \bibinfo {author}
  {\bibfnamefont {C.}~\bibnamefont {Bechinger}}, \ and\ \bibinfo {author}
  {\bibfnamefont {T.}~\bibnamefont {Speck}},\ }\bibfield  {title} {\enquote
  {\bibinfo {title} {Dynamical clustering and phase separation in suspensions
  of self-propelled colloidal particles},}\ }\href {\doibase
  10.1103/PhysRevLett.110.238301} {\bibfield  {journal} {\bibinfo  {journal}
  {Phys. Rev. Lett.}\ }\textbf {\bibinfo {volume} {110}},\ \bibinfo {pages}
  {238301} (\bibinfo {year} {2013})}\BibitemShut {NoStop}%
\bibitem [{\citenamefont {Cates}\ and\ \citenamefont
  {Tailleur}(2015)}]{Cates15}%
  \BibitemOpen
  \bibfield  {author} {\bibinfo {author} {\bibfnamefont {M.~E.}\ \bibnamefont
  {Cates}}\ and\ \bibinfo {author} {\bibfnamefont {J.}~\bibnamefont
  {Tailleur}},\ }\bibfield  {title} {\enquote {\bibinfo {title}
  {Motility-induced phase separation},}\ }\href {\doibase
  10.1146/annurev-conmatphys-031214-014710} {\bibfield  {journal} {\bibinfo
  {journal} {Annual Review of Condensed Matter Physics}\ }\textbf {\bibinfo
  {volume} {6}},\ \bibinfo {pages} {219--244} (\bibinfo {year}
  {2015})}\BibitemShut {NoStop}%
\bibitem [{\citenamefont {Fily}\ \emph {et~al.}(2014)\citenamefont {Fily},
  \citenamefont {Baskaran},\ and\ \citenamefont {Hagan}}]{Fily14a}%
  \BibitemOpen
  \bibfield  {author} {\bibinfo {author} {\bibfnamefont {Y.}~\bibnamefont
  {Fily}}, \bibinfo {author} {\bibfnamefont {A.}~\bibnamefont {Baskaran}}, \
  and\ \bibinfo {author} {\bibfnamefont {M.~F.}\ \bibnamefont {Hagan}},\
  }\bibfield  {title} {\enquote {\bibinfo {title} {Dynamics of self-propelled
  particles under strong confinement},}\ }\href {\doibase 10.1039/c4sm00975d}
  {\bibfield  {journal} {\bibinfo  {journal} {Soft Matter}\ }\textbf {\bibinfo
  {volume} {10}},\ \bibinfo {pages} {5609--5617} (\bibinfo {year}
  {2014})}\BibitemShut {NoStop}%
\bibitem [{\citenamefont {Sartori}\ \emph {et~al.}(2018)\citenamefont
  {Sartori}, \citenamefont {Chiarello}, \citenamefont {Jayaswal}, \citenamefont
  {Pierno}, \citenamefont {Mistura}, \citenamefont {Brun}, \citenamefont
  {Tiribocchi},\ and\ \citenamefont {Orlandini}}]{Sartori18}%
  \BibitemOpen
  \bibfield  {author} {\bibinfo {author} {\bibfnamefont {P.}~\bibnamefont
  {Sartori}}, \bibinfo {author} {\bibfnamefont {E.}~\bibnamefont {Chiarello}},
  \bibinfo {author} {\bibfnamefont {G.}~\bibnamefont {Jayaswal}}, \bibinfo
  {author} {\bibfnamefont {M.}~\bibnamefont {Pierno}}, \bibinfo {author}
  {\bibfnamefont {G.}~\bibnamefont {Mistura}}, \bibinfo {author} {\bibfnamefont
  {P.}~\bibnamefont {Brun}}, \bibinfo {author} {\bibfnamefont {A.}~\bibnamefont
  {Tiribocchi}}, \ and\ \bibinfo {author} {\bibfnamefont {E.}~\bibnamefont
  {Orlandini}},\ }\bibfield  {title} {\enquote {\bibinfo {title} {Wall
  accumulation of bacteria with different motility patterns},}\ }\href
  {\doibase 10.1103/PhysRevE.97.022610} {\bibfield  {journal} {\bibinfo
  {journal} {Phys. Rev. E}\ }\textbf {\bibinfo {volume} {97}},\ \bibinfo
  {pages} {022610} (\bibinfo {year} {2018})}\BibitemShut {NoStop}%
\bibitem [{\citenamefont {Speck}(2020)}]{Speck20}%
  \BibitemOpen
  \bibfield  {author} {\bibinfo {author} {\bibfnamefont {T.}~\bibnamefont
  {Speck}},\ }\bibfield  {title} {\enquote {\bibinfo {title} {Collective forces
  in scalar active matter},}\ }\href {\doibase 10.1039/DoSM00176G} {\bibfield
  {journal} {\bibinfo  {journal} {Soft Matter}\ }\textbf {\bibinfo {volume}
  {16}},\ \bibinfo {pages} {2652} (\bibinfo {year} {2020})}\BibitemShut
  {NoStop}%
\bibitem [{\citenamefont {Das}\ \emph {et~al.}(2020)\citenamefont {Das},
  \citenamefont {Ghosh},\ and\ \citenamefont {Chelakkot}}]{Das20}%
  \BibitemOpen
  \bibfield  {author} {\bibinfo {author} {\bibfnamefont {S.}~\bibnamefont
  {Das}}, \bibinfo {author} {\bibfnamefont {S.}~\bibnamefont {Ghosh}}, \ and\
  \bibinfo {author} {\bibfnamefont {R.}~\bibnamefont {Chelakkot}},\ }\bibfield
  {title} {\enquote {\bibinfo {title} {Aggregate morphology of active
  {B}rownian particles on porous, circular walls},}\ }\href {\doibase
  10.1103/PhysRevE.102.032619} {\bibfield  {journal} {\bibinfo  {journal}
  {Phys. Rev. E}\ }\textbf {\bibinfo {volume} {102}},\ \bibinfo {pages}
  {032619} (\bibinfo {year} {2020})}\BibitemShut {NoStop}%
\bibitem [{\citenamefont {Fazli}\ and\ \citenamefont {Naji}(2021)}]{Fazli21}%
  \BibitemOpen
  \bibfield  {author} {\bibinfo {author} {\bibfnamefont {Z.}~\bibnamefont
  {Fazli}}\ and\ \bibinfo {author} {\bibfnamefont {A.}~\bibnamefont {Naji}},\
  }\bibfield  {title} {\enquote {\bibinfo {title} {Active particles with polar
  alignment in ring-shaped confinement},}\ }\href {\doibase
  10.1103/PhysRevE.103.022601} {\bibfield  {journal} {\bibinfo  {journal}
  {Phys. Rev. E}\ }\textbf {\bibinfo {volume} {103}},\ \bibinfo {pages}
  {022601} (\bibinfo {year} {2021})}\BibitemShut {NoStop}%
\bibitem [{\citenamefont {Ray}\ \emph {et~al.}(2014)\citenamefont {Ray},
  \citenamefont {Reichhardt},\ and\ \citenamefont {Reichhardt}}]{Ray14}%
  \BibitemOpen
  \bibfield  {author} {\bibinfo {author} {\bibfnamefont {D.}~\bibnamefont
  {Ray}}, \bibinfo {author} {\bibfnamefont {C.}~\bibnamefont {Reichhardt}}, \
  and\ \bibinfo {author} {\bibfnamefont {C.~J.~Olson}\ \bibnamefont
  {Reichhardt}},\ }\bibfield  {title} {\enquote {\bibinfo {title} {Casimir
  effect in active matter systems},}\ }\href {\doibase
  10.1103/PhysRevE.90.013019} {\bibfield  {journal} {\bibinfo  {journal} {Phys.
  Rev. E}\ }\textbf {\bibinfo {volume} {90}},\ \bibinfo {pages} {013019}
  (\bibinfo {year} {2014})}\BibitemShut {NoStop}%
\bibitem [{\citenamefont {Ni}\ \emph {et~al.}(2015)\citenamefont {Ni},
  \citenamefont {Cohen~Stuart},\ and\ \citenamefont {Bolhuis}}]{Ni15}%
  \BibitemOpen
  \bibfield  {author} {\bibinfo {author} {\bibfnamefont {R.}~\bibnamefont
  {Ni}}, \bibinfo {author} {\bibfnamefont {M.~A.}\ \bibnamefont
  {Cohen~Stuart}}, \ and\ \bibinfo {author} {\bibfnamefont {P.~G.}\
  \bibnamefont {Bolhuis}},\ }\bibfield  {title} {\enquote {\bibinfo {title}
  {Tunable long range forces mediated by self-propelled colloidal hard
  spheres},}\ }\href {\doibase 10.1103/PhysRevLett.114.018302} {\bibfield
  {journal} {\bibinfo  {journal} {Phys. Rev. Lett.}\ }\textbf {\bibinfo
  {volume} {114}},\ \bibinfo {pages} {018302} (\bibinfo {year}
  {2015})}\BibitemShut {NoStop}%
\bibitem [{\citenamefont {Leite}\ \emph {et~al.}(2016)\citenamefont {Leite},
  \citenamefont {Lucena}, \citenamefont {Potiguar},\ and\ \citenamefont
  {Ferreira}}]{Leite16}%
  \BibitemOpen
  \bibfield  {author} {\bibinfo {author} {\bibfnamefont {L.~R.}\ \bibnamefont
  {Leite}}, \bibinfo {author} {\bibfnamefont {D.}~\bibnamefont {Lucena}},
  \bibinfo {author} {\bibfnamefont {F.~Q.}\ \bibnamefont {Potiguar}}, \ and\
  \bibinfo {author} {\bibfnamefont {W.~P.}\ \bibnamefont {Ferreira}},\
  }\bibfield  {title} {\enquote {\bibinfo {title} {Depletion forces on circular
  and elliptical obstacles induced by active matter},}\ }\href {\doibase
  10.1103/PhysRevE.94.062602} {\bibfield  {journal} {\bibinfo  {journal} {Phys.
  Rev. E}\ }\textbf {\bibinfo {volume} {94}},\ \bibinfo {pages} {062602}
  (\bibinfo {year} {2016})}\BibitemShut {NoStop}%
\bibitem [{\citenamefont {Mallory}\ \emph {et~al.}(2018)\citenamefont
  {Mallory}, \citenamefont {Valeriani},\ and\ \citenamefont
  {Cacciuto}}]{Mallory18}%
  \BibitemOpen
  \bibfield  {author} {\bibinfo {author} {\bibfnamefont {S.~A.}\ \bibnamefont
  {Mallory}}, \bibinfo {author} {\bibfnamefont {C.}~\bibnamefont {Valeriani}},
  \ and\ \bibinfo {author} {\bibfnamefont {A.}~\bibnamefont {Cacciuto}},\
  }\bibfield  {title} {\enquote {\bibinfo {title} {An active approach to
  colloidal self-assembly},}\ }\href {\doibase
  10.1146/annurev-physchem-050317-021237} {\bibfield  {journal} {\bibinfo
  {journal} {Ann. Rev. Phys. Chem.}\ }\textbf {\bibinfo {volume} {69}},\
  \bibinfo {pages} {59} (\bibinfo {year} {2018})}\BibitemShut {NoStop}%
\bibitem [{\citenamefont {Kjeldbjerg}\ and\ \citenamefont
  {Brady}(2021)}]{Kjeldbjerg21}%
  \BibitemOpen
  \bibfield  {author} {\bibinfo {author} {\bibfnamefont {C.~M.}\ \bibnamefont
  {Kjeldbjerg}}\ and\ \bibinfo {author} {\bibfnamefont {J.~F.}\ \bibnamefont
  {Brady}},\ }\bibfield  {title} {\enquote {\bibinfo {title} {Theory for the
  {C}asimir effect and the partitioning of active matter},}\ }\href {\doibase
  10.1039/D0SM01797C} {\bibfield  {journal} {\bibinfo  {journal} {Soft Matter}\
  }\textbf {\bibinfo {volume} {17}},\ \bibinfo {pages} {523} (\bibinfo {year}
  {2021})}\BibitemShut {NoStop}%
\bibitem [{\citenamefont {Galajda}\ \emph {et~al.}(2007)\citenamefont
  {Galajda}, \citenamefont {Keymer}, \citenamefont {Chaikin},\ and\
  \citenamefont {Austin}}]{Galajda07}%
  \BibitemOpen
  \bibfield  {author} {\bibinfo {author} {\bibfnamefont {P.}~\bibnamefont
  {Galajda}}, \bibinfo {author} {\bibfnamefont {J.}~\bibnamefont {Keymer}},
  \bibinfo {author} {\bibfnamefont {P.}~\bibnamefont {Chaikin}}, \ and\
  \bibinfo {author} {\bibfnamefont {R.}~\bibnamefont {Austin}},\ }\bibfield
  {title} {\enquote {\bibinfo {title} {A wall of funnels concentrates swimming
  bacteria},}\ }\href {\doibase 10.1128/JB.01033-07} {\bibfield  {journal}
  {\bibinfo  {journal} {J. Bacteriol.}\ }\textbf {\bibinfo {volume} {189}},\
  \bibinfo {pages} {8704--8707} (\bibinfo {year} {2007})}\BibitemShut {NoStop}%
\bibitem [{\citenamefont {Tailleur}\ and\ \citenamefont
  {Cates}(2009)}]{Tailleur09}%
  \BibitemOpen
  \bibfield  {author} {\bibinfo {author} {\bibfnamefont {J.}~\bibnamefont
  {Tailleur}}\ and\ \bibinfo {author} {\bibfnamefont {M.~E.}\ \bibnamefont
  {Cates}},\ }\bibfield  {title} {\enquote {\bibinfo {title} {Sedimentation,
  trapping, and rectification of dilute bacteria},}\ }\href {\doibase
  10.1209/0295-5075/86/60002} {\bibfield  {journal} {\bibinfo  {journal} {EPL}\
  }\textbf {\bibinfo {volume} {86}},\ \bibinfo {pages} {60002} (\bibinfo {year}
  {2009})}\BibitemShut {NoStop}%
\bibitem [{\citenamefont {Ai}(2016)}]{Ai16}%
  \BibitemOpen
  \bibfield  {author} {\bibinfo {author} {\bibfnamefont {B.}~\bibnamefont
  {Ai}},\ }\bibfield  {title} {\enquote {\bibinfo {title} {Ratchet transport
  powered by chiral active particles},}\ }\href {\doibase 10.1038/srep18740}
  {\bibfield  {journal} {\bibinfo  {journal} {Sci. Rep.}\ }\textbf {\bibinfo
  {volume} {6}},\ \bibinfo {pages} {18740} (\bibinfo {year}
  {2016})}\BibitemShut {NoStop}%
\bibitem [{\citenamefont {Reichhardt}\ and\ \citenamefont
  {Reichhardt}(2017{\natexlab{a}})}]{Reichhardt17a}%
  \BibitemOpen
  \bibfield  {author} {\bibinfo {author} {\bibfnamefont {C.~J.~Olson}\
  \bibnamefont {Reichhardt}}\ and\ \bibinfo {author} {\bibfnamefont
  {C.}~\bibnamefont {Reichhardt}},\ }\bibfield  {title} {\enquote {\bibinfo
  {title} {Ratchet effects in active matter systems},}\ }\href {\doibase
  10.1146/annurev-conmatphys-031016-025522} {\bibfield  {journal} {\bibinfo
  {journal} {Ann. Rev. Condens. Matter Phys.}\ }\textbf {\bibinfo {volume}
  {8}},\ \bibinfo {pages} {51--75} (\bibinfo {year}
  {2017}{\natexlab{a}})}\BibitemShut {NoStop}%
\bibitem [{\citenamefont {Borba}\ \emph {et~al.}(2020)\citenamefont {Borba},
  \citenamefont {Domingos}, \citenamefont {Moraes}, \citenamefont {Potiguar},\
  and\ \citenamefont {Ferreira}}]{Borba20}%
  \BibitemOpen
  \bibfield  {author} {\bibinfo {author} {\bibfnamefont {A.~D.}\ \bibnamefont
  {Borba}}, \bibinfo {author} {\bibfnamefont {Jorge L.~C.}\ \bibnamefont
  {Domingos}}, \bibinfo {author} {\bibfnamefont {E.~C.~B.}\ \bibnamefont
  {Moraes}}, \bibinfo {author} {\bibfnamefont {F.~Q.}\ \bibnamefont
  {Potiguar}}, \ and\ \bibinfo {author} {\bibfnamefont {W.~P.}\ \bibnamefont
  {Ferreira}},\ }\bibfield  {title} {\enquote {\bibinfo {title} {Controlling
  the transport of active matter in disordered lattices of asymmetrical
  obstacles},}\ }\href {\doibase 10.1103/PhysRevE.101.022601} {\bibfield
  {journal} {\bibinfo  {journal} {Phys. Rev. E}\ }\textbf {\bibinfo {volume}
  {101}},\ \bibinfo {pages} {022601} (\bibinfo {year} {2020})}\BibitemShut
  {NoStop}%
\bibitem [{\citenamefont {Khatami}\ \emph {et~al.}(2016)\citenamefont
  {Khatami}, \citenamefont {Wolff}, \citenamefont {Pohl}, \citenamefont
  {Ejtehadi},\ and\ \citenamefont {Stark}}]{Khatami16}%
  \BibitemOpen
  \bibfield  {author} {\bibinfo {author} {\bibfnamefont {M.}~\bibnamefont
  {Khatami}}, \bibinfo {author} {\bibfnamefont {K.}~\bibnamefont {Wolff}},
  \bibinfo {author} {\bibfnamefont {O.}~\bibnamefont {Pohl}}, \bibinfo {author}
  {\bibfnamefont {M.~R.}\ \bibnamefont {Ejtehadi}}, \ and\ \bibinfo {author}
  {\bibfnamefont {H.}~\bibnamefont {Stark}},\ }\bibfield  {title} {\enquote
  {\bibinfo {title} {Active {B}rownian particles and run-and-tumble particles
  separate inside a maze},}\ }\href {\doibase 10.1038/srep37670} {\bibfield
  {journal} {\bibinfo  {journal} {Sci. Rep.}\ }\textbf {\bibinfo {volume}
  {6}},\ \bibinfo {pages} {37670} (\bibinfo {year} {2016})}\BibitemShut
  {NoStop}%
\bibitem [{\citenamefont {Yang}\ and\ \citenamefont {Bevan}(2018)}]{Yang18}%
  \BibitemOpen
  \bibfield  {author} {\bibinfo {author} {\bibfnamefont {Y.}~\bibnamefont
  {Yang}}\ and\ \bibinfo {author} {\bibfnamefont {M.~A.}\ \bibnamefont
  {Bevan}},\ }\bibfield  {title} {\enquote {\bibinfo {title} {Optimal
  navigation of self-propelled colloids},}\ }\href {\doibase
  10.1021/acsnano.8b05371} {\bibfield  {journal} {\bibinfo  {journal} {ACS
  Nano}\ }\textbf {\bibinfo {volume} {12}},\ \bibinfo {pages} {10712} (\bibinfo
  {year} {2018})}\BibitemShut {NoStop}%
\bibitem [{\citenamefont {Reichhardt}\ and\ \citenamefont
  {Reichhardt}(2018{\natexlab{a}})}]{Reichhardt18c}%
  \BibitemOpen
  \bibfield  {author} {\bibinfo {author} {\bibfnamefont {C.}~\bibnamefont
  {Reichhardt}}\ and\ \bibinfo {author} {\bibfnamefont {C.~J.~O.}\ \bibnamefont
  {Reichhardt}},\ }\bibfield  {title} {\enquote {\bibinfo {title} {Clogging and
  depinning of ballistic active matter systems in disordered media},}\ }\href
  {\doibase 10.1103/PhysRevE.97.052613} {\bibfield  {journal} {\bibinfo
  {journal} {Phys. Rev. E}\ }\textbf {\bibinfo {volume} {97}},\ \bibinfo
  {pages} {052613} (\bibinfo {year} {2018}{\natexlab{a}})}\BibitemShut
  {NoStop}%
\bibitem [{\citenamefont {Caprini}\ \emph {et~al.}(2020)\citenamefont
  {Caprini}, \citenamefont {Cecconi}, \citenamefont {Maggi},\ and\
  \citenamefont {Marini Bettolo~Marconi}}]{Caprini20}%
  \BibitemOpen
  \bibfield  {author} {\bibinfo {author} {\bibfnamefont {L.}~\bibnamefont
  {Caprini}}, \bibinfo {author} {\bibfnamefont {F.}~\bibnamefont {Cecconi}},
  \bibinfo {author} {\bibfnamefont {C.}~\bibnamefont {Maggi}}, \ and\ \bibinfo
  {author} {\bibfnamefont {U.}~\bibnamefont {Marini Bettolo~Marconi}},\
  }\bibfield  {title} {\enquote {\bibinfo {title} {Activity-controlled clogging
  and unclogging of microchannels},}\ }\href {\doibase
  10.1103/PhysRevResearch.2.043359} {\bibfield  {journal} {\bibinfo  {journal}
  {Phys. Rev. Research}\ }\textbf {\bibinfo {volume} {2}},\ \bibinfo {pages}
  {043359} (\bibinfo {year} {2020})}\BibitemShut {NoStop}%
\bibitem [{\citenamefont {Shi}\ \emph {et~al.}(2020)\citenamefont {Shi},
  \citenamefont {Li}, \citenamefont {Feng}, \citenamefont {Tian},\ and\
  \citenamefont {Chen}}]{Shi20}%
  \BibitemOpen
  \bibfield  {author} {\bibinfo {author} {\bibfnamefont {S.}~\bibnamefont
  {Shi}}, \bibinfo {author} {\bibfnamefont {H.}~\bibnamefont {Li}}, \bibinfo
  {author} {\bibfnamefont {G.}~\bibnamefont {Feng}}, \bibinfo {author}
  {\bibfnamefont {W.}~\bibnamefont {Tian}}, \ and\ \bibinfo {author}
  {\bibfnamefont {K}~\bibnamefont {Chen}},\ }\bibfield  {title} {\enquote
  {\bibinfo {title} {Transport of self-propelled particles across a porous
  medium: trapping, clogging, and the {M}atthew effect},}\ }\href {\doibase
  10.1039/D0CP01923B} {\bibfield  {journal} {\bibinfo  {journal} {Phys. Chem.
  Chem. Phys.}\ }\textbf {\bibinfo {volume} {22}},\ \bibinfo {pages} {14052}
  (\bibinfo {year} {2020})}\BibitemShut {NoStop}%
\bibitem [{\citenamefont {Reichhardt}\ and\ \citenamefont
  {Olson~Reichhardt}(2014)}]{Reichhardt14}%
  \BibitemOpen
  \bibfield  {author} {\bibinfo {author} {\bibfnamefont {C.}~\bibnamefont
  {Reichhardt}}\ and\ \bibinfo {author} {\bibfnamefont {C.~J.}\ \bibnamefont
  {Olson~Reichhardt}},\ }\bibfield  {title} {\enquote {\bibinfo {title} {Active
  matter transport and jamming on disordered landscapes},}\ }\href {\doibase
  10.1103/PhysRevE.90.012701} {\bibfield  {journal} {\bibinfo  {journal} {Phys.
  Rev. E}\ }\textbf {\bibinfo {volume} {90}},\ \bibinfo {pages} {012701}
  (\bibinfo {year} {2014})}\BibitemShut {NoStop}%
\bibitem [{\citenamefont {Zeitz}\ \emph {et~al.}(2017)\citenamefont {Zeitz},
  \citenamefont {Wolff},\ and\ \citenamefont {Stark}}]{Zeitz17}%
  \BibitemOpen
  \bibfield  {author} {\bibinfo {author} {\bibfnamefont {M.}~\bibnamefont
  {Zeitz}}, \bibinfo {author} {\bibfnamefont {K.}~\bibnamefont {Wolff}}, \ and\
  \bibinfo {author} {\bibfnamefont {H.}~\bibnamefont {Stark}},\ }\bibfield
  {title} {\enquote {\bibinfo {title} {Active {B}rownian particles moving in a
  random {L}orentz gas},}\ }\href {\doibase 10.1140/epje/i2017-11510-0}
  {\bibfield  {journal} {\bibinfo  {journal} {Eur. Phys. J. E}\ }\textbf
  {\bibinfo {volume} {40}},\ \bibinfo {pages} {23} (\bibinfo {year}
  {2017})}\BibitemShut {NoStop}%
\bibitem [{\citenamefont {S{\' a}ndor}\ \emph {et~al.}(2017)\citenamefont {S{\'
  a}ndor}, \citenamefont {Lib{\' a}l}, \citenamefont {Reichhardt},\ and\
  \citenamefont {Reichhardt}}]{Sandor17b}%
  \BibitemOpen
  \bibfield  {author} {\bibinfo {author} {\bibfnamefont {Cs.}\ \bibnamefont
  {S{\' a}ndor}}, \bibinfo {author} {\bibfnamefont {A.}~\bibnamefont {Lib{\'
  a}l}}, \bibinfo {author} {\bibfnamefont {C.}~\bibnamefont {Reichhardt}}, \
  and\ \bibinfo {author} {\bibfnamefont {C.~J.~Olson}\ \bibnamefont
  {Reichhardt}},\ }\bibfield  {title} {\enquote {\bibinfo {title} {Dewetting
  and spreading transitions for active matter on random pinning substrates},}\
  }\href {\doibase 10.1063/1.4983344} {\bibfield  {journal} {\bibinfo
  {journal} {J. Chem. Phys.}\ }\textbf {\bibinfo {volume} {146}},\ \bibinfo
  {pages} {204903} (\bibinfo {year} {2017})}\BibitemShut {NoStop}%
\bibitem [{\citenamefont {Morin}\ \emph
  {et~al.}(2017{\natexlab{a}})\citenamefont {Morin}, \citenamefont
  {Lopes~Cardozo}, \citenamefont {Chikkadi},\ and\ \citenamefont
  {Bartolo}}]{Morin17a}%
  \BibitemOpen
  \bibfield  {author} {\bibinfo {author} {\bibfnamefont {A.}~\bibnamefont
  {Morin}}, \bibinfo {author} {\bibfnamefont {D.}~\bibnamefont
  {Lopes~Cardozo}}, \bibinfo {author} {\bibfnamefont {V.}~\bibnamefont
  {Chikkadi}}, \ and\ \bibinfo {author} {\bibfnamefont {D.}~\bibnamefont
  {Bartolo}},\ }\bibfield  {title} {\enquote {\bibinfo {title} {Diffusion,
  subdiffusion, and localization of active colloids in random post lattices},}\
  }\href {\doibase 10.1103/PhysRevE.96.042611} {\bibfield  {journal} {\bibinfo
  {journal} {Phys. Rev. E}\ }\textbf {\bibinfo {volume} {96}},\ \bibinfo
  {pages} {042611} (\bibinfo {year} {2017}{\natexlab{a}})}\BibitemShut
  {NoStop}%
\bibitem [{\citenamefont {Chepizhko}\ \emph {et~al.}(2013)\citenamefont
  {Chepizhko}, \citenamefont {Altmann},\ and\ \citenamefont
  {Peruani}}]{Chepizhko13}%
  \BibitemOpen
  \bibfield  {author} {\bibinfo {author} {\bibfnamefont {O.}~\bibnamefont
  {Chepizhko}}, \bibinfo {author} {\bibfnamefont {E.~G.}\ \bibnamefont
  {Altmann}}, \ and\ \bibinfo {author} {\bibfnamefont {F.}~\bibnamefont
  {Peruani}},\ }\bibfield  {title} {\enquote {\bibinfo {title} {Optimal noise
  maximizes collective motion in heterogeneous media},}\ }\href {\doibase
  10.1103/PhysRevLett.110.238101} {\bibfield  {journal} {\bibinfo  {journal}
  {Phys. Rev. Lett.}\ }\textbf {\bibinfo {volume} {110}},\ \bibinfo {pages}
  {238101} (\bibinfo {year} {2013})}\BibitemShut {NoStop}%
\bibitem [{\citenamefont {Bertrand}\ \emph {et~al.}(2018)\citenamefont
  {Bertrand}, \citenamefont {Zhao}, \citenamefont {B\'enichou}, \citenamefont
  {Tailleur},\ and\ \citenamefont {Voituriez}}]{Bertrand18}%
  \BibitemOpen
  \bibfield  {author} {\bibinfo {author} {\bibfnamefont {T.}~\bibnamefont
  {Bertrand}}, \bibinfo {author} {\bibfnamefont {Y.}~\bibnamefont {Zhao}},
  \bibinfo {author} {\bibfnamefont {O.}~\bibnamefont {B\'enichou}}, \bibinfo
  {author} {\bibfnamefont {J.}~\bibnamefont {Tailleur}}, \ and\ \bibinfo
  {author} {\bibfnamefont {R.}~\bibnamefont {Voituriez}},\ }\bibfield  {title}
  {\enquote {\bibinfo {title} {Optimized diffusion of run-and-tumble particles
  in crowded environments},}\ }\href {\doibase 10.1103/PhysRevLett.120.198103}
  {\bibfield  {journal} {\bibinfo  {journal} {Phys. Rev. Lett.}\ }\textbf
  {\bibinfo {volume} {120}},\ \bibinfo {pages} {198103} (\bibinfo {year}
  {2018})}\BibitemShut {NoStop}%
\bibitem [{\citenamefont {Chepizhko}\ and\ \citenamefont
  {Franosch}(2019)}]{Chepizhko19}%
  \BibitemOpen
  \bibfield  {author} {\bibinfo {author} {\bibfnamefont {O.}~\bibnamefont
  {Chepizhko}}\ and\ \bibinfo {author} {\bibfnamefont {T.}~\bibnamefont
  {Franosch}},\ }\bibfield  {title} {\enquote {\bibinfo {title} {Ideal circle
  microswimmers in crowded media},}\ }\href {\doibase 10.1039/c8sm02030b}
  {\bibfield  {journal} {\bibinfo  {journal} {Soft Matter}\ }\textbf {\bibinfo
  {volume} {15}},\ \bibinfo {pages} {452--461} (\bibinfo {year}
  {2019})}\BibitemShut {NoStop}%
\bibitem [{\citenamefont {Bhattacharjee}\ and\ \citenamefont
  {Datta}(2019)}]{Bhattacharjee19}%
  \BibitemOpen
  \bibfield  {author} {\bibinfo {author} {\bibfnamefont {T.}~\bibnamefont
  {Bhattacharjee}}\ and\ \bibinfo {author} {\bibfnamefont {S.~S.}\ \bibnamefont
  {Datta}},\ }\bibfield  {title} {\enquote {\bibinfo {title} {Confinement and
  activity regulate bacterial motion in porous media},}\ }\href {\doibase
  10.1039/C9SM01735F} {\bibfield  {journal} {\bibinfo  {journal} {Soft Matter}\
  }\textbf {\bibinfo {volume} {15}},\ \bibinfo {pages} {9920} (\bibinfo {year}
  {2019})}\BibitemShut {NoStop}%
\bibitem [{\citenamefont {Breoni}\ \emph {et~al.}(2020)\citenamefont {Breoni},
  \citenamefont {Schmiedeberg},\ and\ \citenamefont {L\"owen}}]{Breoni20}%
  \BibitemOpen
  \bibfield  {author} {\bibinfo {author} {\bibfnamefont {D.}~\bibnamefont
  {Breoni}}, \bibinfo {author} {\bibfnamefont {M.}~\bibnamefont
  {Schmiedeberg}}, \ and\ \bibinfo {author} {\bibfnamefont {H.}~\bibnamefont
  {L\"owen}},\ }\bibfield  {title} {\enquote {\bibinfo {title} {Active
  {B}rownian and inertial particles in disordered environments: Short-time
  expansion of the mean-square displacement},}\ }\href {\doibase
  10.1103/PhysRevE.102.062604} {\bibfield  {journal} {\bibinfo  {journal}
  {Phys. Rev. E}\ }\textbf {\bibinfo {volume} {102}},\ \bibinfo {pages}
  {062604} (\bibinfo {year} {2020})}\BibitemShut {NoStop}%
\bibitem [{\citenamefont {S\'andor}\ \emph {et~al.}(2017)\citenamefont
  {S\'andor}, \citenamefont {Lib\'al}, \citenamefont {Reichhardt},\ and\
  \citenamefont {Olson~Reichhardt}}]{Sandor17a}%
  \BibitemOpen
  \bibfield  {author} {\bibinfo {author} {\bibfnamefont {Cs.}\ \bibnamefont
  {S\'andor}}, \bibinfo {author} {\bibfnamefont {A.}~\bibnamefont {Lib\'al}},
  \bibinfo {author} {\bibfnamefont {C.}~\bibnamefont {Reichhardt}}, \ and\
  \bibinfo {author} {\bibfnamefont {C.~J.}\ \bibnamefont {Olson~Reichhardt}},\
  }\bibfield  {title} {\enquote {\bibinfo {title} {Dynamic phases of active
  matter systems with quenched disorder},}\ }\href {\doibase
  10.1103/PhysRevE.95.032606} {\bibfield  {journal} {\bibinfo  {journal} {Phys.
  Rev. E}\ }\textbf {\bibinfo {volume} {95}},\ \bibinfo {pages} {032606}
  (\bibinfo {year} {2017})}\BibitemShut {NoStop}%
\bibitem [{\citenamefont {Reichhardt}\ and\ \citenamefont
  {Reichhardt}(2018{\natexlab{b}})}]{Reichhardt18a}%
  \BibitemOpen
  \bibfield  {author} {\bibinfo {author} {\bibfnamefont {C.~J.~O.}\
  \bibnamefont {Reichhardt}}\ and\ \bibinfo {author} {\bibfnamefont
  {C.}~\bibnamefont {Reichhardt}},\ }\bibfield  {title} {\enquote {\bibinfo
  {title} {Avalanche dynamics for active matter in heterogeneous media},}\
  }\href {\doibase 10.1088/1367-2630/aaa392} {\bibfield  {journal} {\bibinfo
  {journal} {New J. Phys.}\ }\textbf {\bibinfo {volume} {20}},\ \bibinfo
  {pages} {025002} (\bibinfo {year} {2018}{\natexlab{b}})}\BibitemShut
  {NoStop}%
\bibitem [{\citenamefont {Morin}\ \emph
  {et~al.}(2017{\natexlab{b}})\citenamefont {Morin}, \citenamefont
  {Desreumaux}, \citenamefont {Caussin},\ and\ \citenamefont
  {Bartolo}}]{Morin17}%
  \BibitemOpen
  \bibfield  {author} {\bibinfo {author} {\bibfnamefont {A.}~\bibnamefont
  {Morin}}, \bibinfo {author} {\bibfnamefont {N.}~\bibnamefont {Desreumaux}},
  \bibinfo {author} {\bibfnamefont {J.-B.}\ \bibnamefont {Caussin}}, \ and\
  \bibinfo {author} {\bibfnamefont {D.}~\bibnamefont {Bartolo}},\ }\bibfield
  {title} {\enquote {\bibinfo {title} {Distortion and destruction of colloidal
  flocks in disordered environments},}\ }\href {\doibase 10.1038/nphys3903}
  {\bibfield  {journal} {\bibinfo  {journal} {Nature Phys.}\ }\textbf {\bibinfo
  {volume} {13}},\ \bibinfo {pages} {63--67} (\bibinfo {year}
  {2017}{\natexlab{b}})}\BibitemShut {NoStop}%
\bibitem [{\citenamefont {Bijnens}\ and\ \citenamefont
  {Maes}(2021)}]{Bijnens21}%
  \BibitemOpen
  \bibfield  {author} {\bibinfo {author} {\bibfnamefont {B.}~\bibnamefont
  {Bijnens}}\ and\ \bibinfo {author} {\bibfnamefont {C.}~\bibnamefont {Maes}},\
  }\bibfield  {title} {\enquote {\bibinfo {title} {Pushing run-and-tumble
  particles through a rugged channel},}\ }\href {\doibase
  10.1088/1742-5468/abe29e} {\bibfield  {journal} {\bibinfo  {journal} {J.
  Stat. Mech.: Theor. Exp.}\ }\textbf {\bibinfo {volume} {2021}},\ \bibinfo
  {pages} {033206} (\bibinfo {year} {2021})}\BibitemShut {NoStop}%
\bibitem [{\citenamefont {Chardac}\ \emph {et~al.}(2021)\citenamefont
  {Chardac}, \citenamefont {Shankar}, \citenamefont {Marchetti},\ and\
  \citenamefont {Bartolo}}]{Chardac21}%
  \BibitemOpen
  \bibfield  {author} {\bibinfo {author} {\bibfnamefont {A.}~\bibnamefont
  {Chardac}}, \bibinfo {author} {\bibfnamefont {S.}~\bibnamefont {Shankar}},
  \bibinfo {author} {\bibfnamefont {M.~C.}\ \bibnamefont {Marchetti}}, \ and\
  \bibinfo {author} {\bibfnamefont {D.}~\bibnamefont {Bartolo}},\ }\bibfield
  {title} {\enquote {\bibinfo {title} {Emergence of dynamic vortex glasses in
  disordered polar active fluids},}\ }\href {\doibase 10.1073/pnas.2018218118}
  {\bibfield  {journal} {\bibinfo  {journal} {Proc. Natl. Acad. Sci. (USA)}\
  }\textbf {\bibinfo {volume} {118}},\ \bibinfo {pages} {e2018218118} (\bibinfo
  {year} {2021})}\BibitemShut {NoStop}%
\bibitem [{\citenamefont {Reichhardt}\ and\ \citenamefont
  {Reichhardt}(2014)}]{Reichhardt14a}%
  \BibitemOpen
  \bibfield  {author} {\bibinfo {author} {\bibfnamefont {C.}~\bibnamefont
  {Reichhardt}}\ and\ \bibinfo {author} {\bibfnamefont {C.~J.~Olson}\
  \bibnamefont {Reichhardt}},\ }\bibfield  {title} {\enquote {\bibinfo {title}
  {Absorbing phase transitions and dynamic freezing in running active matter
  systems},}\ }\href {\doibase 10.1039/c4sm01273a} {\bibfield  {journal}
  {\bibinfo  {journal} {Soft Matter}\ }\textbf {\bibinfo {volume} {10}},\
  \bibinfo {pages} {7502--7510} (\bibinfo {year} {2014})}\BibitemShut {NoStop}%
\bibitem [{\citenamefont {Volpe}\ \emph {et~al.}(2011)\citenamefont {Volpe},
  \citenamefont {Buttinoni}, \citenamefont {Vogt}, \citenamefont {K{\"
  u}mmerer},\ and\ \citenamefont {Bechinger}}]{Volpe11}%
  \BibitemOpen
  \bibfield  {author} {\bibinfo {author} {\bibfnamefont {G.}~\bibnamefont
  {Volpe}}, \bibinfo {author} {\bibfnamefont {I.}~\bibnamefont {Buttinoni}},
  \bibinfo {author} {\bibfnamefont {D.}~\bibnamefont {Vogt}}, \bibinfo {author}
  {\bibfnamefont {H.-J.}\ \bibnamefont {K{\" u}mmerer}}, \ and\ \bibinfo
  {author} {\bibfnamefont {C.}~\bibnamefont {Bechinger}},\ }\bibfield  {title}
  {\enquote {\bibinfo {title} {Microswimmers in patterned environments},}\
  }\href {\doibase 10.1039/c1sm05960b} {\bibfield  {journal} {\bibinfo
  {journal} {Soft Matter}\ }\textbf {\bibinfo {volume} {7}},\ \bibinfo {pages}
  {8810--8815} (\bibinfo {year} {2011})}\BibitemShut {NoStop}%
\bibitem [{\citenamefont {Reichhardt}\ and\ \citenamefont
  {Reichhardt}(2020{\natexlab{a}})}]{Reichhardt20}%
  \BibitemOpen
  \bibfield  {author} {\bibinfo {author} {\bibfnamefont {C.}~\bibnamefont
  {Reichhardt}}\ and\ \bibinfo {author} {\bibfnamefont {C.~J.~O.}\ \bibnamefont
  {Reichhardt}},\ }\bibfield  {title} {\enquote {\bibinfo {title} {Directional
  locking effects for active matter particles coupled to a periodic
  substrate},}\ }\href {\doibase 10.1103/PhysRevE.102.042616} {\bibfield
  {journal} {\bibinfo  {journal} {Phys. Rev. E}\ }\textbf {\bibinfo {volume}
  {102}},\ \bibinfo {pages} {042616} (\bibinfo {year}
  {2020}{\natexlab{a}})}\BibitemShut {NoStop}%
\bibitem [{\citenamefont {Brun-Cosme-Bruny}\ \emph {et~al.}(2020)\citenamefont
  {Brun-Cosme-Bruny}, \citenamefont {F\"ortsch}, \citenamefont {Zimmermann},
  \citenamefont {Bertin}, \citenamefont {Peyla},\ and\ \citenamefont
  {Rafa\"{\i}}}]{BrunCosmeBruny20}%
  \BibitemOpen
  \bibfield  {author} {\bibinfo {author} {\bibfnamefont {M.}~\bibnamefont
  {Brun-Cosme-Bruny}}, \bibinfo {author} {\bibfnamefont {A.}~\bibnamefont
  {F\"ortsch}}, \bibinfo {author} {\bibfnamefont {W.}~\bibnamefont
  {Zimmermann}}, \bibinfo {author} {\bibfnamefont {E.}~\bibnamefont {Bertin}},
  \bibinfo {author} {\bibfnamefont {P.}~\bibnamefont {Peyla}}, \ and\ \bibinfo
  {author} {\bibfnamefont {S.}~\bibnamefont {Rafa\"{\i}}},\ }\bibfield  {title}
  {\enquote {\bibinfo {title} {Deflection of phototactic microswimmers through
  obstacle arrays},}\ }\href {\doibase 10.1103/PhysRevFluids.5.093302}
  {\bibfield  {journal} {\bibinfo  {journal} {Phys. Rev. Fluids}\ }\textbf
  {\bibinfo {volume} {5}},\ \bibinfo {pages} {093302} (\bibinfo {year}
  {2020})}\BibitemShut {NoStop}%
\bibitem [{\citenamefont {Pattanayak}\ \emph {et~al.}(2019)\citenamefont
  {Pattanayak}, \citenamefont {Das}, \citenamefont {Kumar},\ and\ \citenamefont
  {Mishra}}]{Pattanayak19}%
  \BibitemOpen
  \bibfield  {author} {\bibinfo {author} {\bibfnamefont {S.}~\bibnamefont
  {Pattanayak}}, \bibinfo {author} {\bibfnamefont {R.}~\bibnamefont {Das}},
  \bibinfo {author} {\bibfnamefont {M.}~\bibnamefont {Kumar}}, \ and\ \bibinfo
  {author} {\bibfnamefont {S.}~\bibnamefont {Mishra}},\ }\bibfield  {title}
  {\enquote {\bibinfo {title} {Enhanced dynamics of active {B}rownian particles
  in periodic obstacle arrays and corrugated channels},}\ }\href {\doibase
  10.1140/epje/i2019-11826-7} {\bibfield  {journal} {\bibinfo  {journal} {Eur.
  Phys. J. E}\ }\textbf {\bibinfo {volume} {42}},\ \bibinfo {pages} {62}
  (\bibinfo {year} {2019})}\BibitemShut {NoStop}%
\bibitem [{\citenamefont {Schakenraad}\ \emph {et~al.}(2020)\citenamefont
  {Schakenraad}, \citenamefont {Ravazzano}, \citenamefont {Sarkar},
  \citenamefont {Wondergem}, \citenamefont {Merks},\ and\ \citenamefont
  {Giomi}}]{Schakenraad20}%
  \BibitemOpen
  \bibfield  {author} {\bibinfo {author} {\bibfnamefont {K.}~\bibnamefont
  {Schakenraad}}, \bibinfo {author} {\bibfnamefont {L.}~\bibnamefont
  {Ravazzano}}, \bibinfo {author} {\bibfnamefont {N.}~\bibnamefont {Sarkar}},
  \bibinfo {author} {\bibfnamefont {J.~A.~J.}\ \bibnamefont {Wondergem}},
  \bibinfo {author} {\bibfnamefont {R.~M.~H.}\ \bibnamefont {Merks}}, \ and\
  \bibinfo {author} {\bibfnamefont {L.}~\bibnamefont {Giomi}},\ }\bibfield
  {title} {\enquote {\bibinfo {title} {Topotaxis of active {B}rownian
  particles},}\ }\href {\doibase 10.1103/PhysRevE.101.032602} {\bibfield
  {journal} {\bibinfo  {journal} {Phys. Rev. E}\ }\textbf {\bibinfo {volume}
  {101}},\ \bibinfo {pages} {032602} (\bibinfo {year} {2020})}\BibitemShut
  {NoStop}%
\bibitem [{\citenamefont {Ribeiro}\ \emph {et~al.}(2020)\citenamefont
  {Ribeiro}, \citenamefont {Ferreira},\ and\ \citenamefont
  {Potiguar}}]{Ribeiro20}%
  \BibitemOpen
  \bibfield  {author} {\bibinfo {author} {\bibfnamefont {H.~E.}\ \bibnamefont
  {Ribeiro}}, \bibinfo {author} {\bibfnamefont {W.~P.}\ \bibnamefont
  {Ferreira}}, \ and\ \bibinfo {author} {\bibfnamefont {Fabr\'{\i}cio~Q.}\
  \bibnamefont {Potiguar}},\ }\bibfield  {title} {\enquote {\bibinfo {title}
  {Trapping and sorting of active matter in a periodic background potential},}\
  }\href {\doibase 10.1103/PhysRevE.101.032126} {\bibfield  {journal} {\bibinfo
   {journal} {Phys. Rev. E}\ }\textbf {\bibinfo {volume} {101}},\ \bibinfo
  {pages} {032126} (\bibinfo {year} {2020})}\BibitemShut {NoStop}%
\bibitem [{\citenamefont {Reichhardt}\ and\ \citenamefont
  {Reichhardt}(2021{\natexlab{a}})}]{Reichhardt21a}%
  \BibitemOpen
  \bibfield  {author} {\bibinfo {author} {\bibfnamefont {C.}~\bibnamefont
  {Reichhardt}}\ and\ \bibinfo {author} {\bibfnamefont {C.~J.~O.}\ \bibnamefont
  {Reichhardt}},\ }\bibfield  {title} {\enquote {\bibinfo {title} {Clogging,
  dynamics, and reentrant fluid for active matter on periodic substrates},}\
  }\href {\doibase 10.1103/PhysRevE.103.062603} {\bibfield  {journal} {\bibinfo
   {journal} {Phys. Rev. E}\ }\textbf {\bibinfo {volume} {103}},\ \bibinfo
  {pages} {062603} (\bibinfo {year} {2021}{\natexlab{a}})}\BibitemShut
  {NoStop}%
\bibitem [{\citenamefont {Reinken}\ \emph {et~al.}(2020)\citenamefont
  {Reinken}, \citenamefont {Nishiguchi}, \citenamefont {Heidenreich},
  \citenamefont {Sokolov}, \citenamefont {B{\" a}r}, \citenamefont {Klapp},\
  and\ \citenamefont {Aranson}}]{Reinken20}%
  \BibitemOpen
  \bibfield  {author} {\bibinfo {author} {\bibfnamefont {H.}~\bibnamefont
  {Reinken}}, \bibinfo {author} {\bibfnamefont {D.}~\bibnamefont {Nishiguchi}},
  \bibinfo {author} {\bibfnamefont {S.}~\bibnamefont {Heidenreich}}, \bibinfo
  {author} {\bibfnamefont {A.}~\bibnamefont {Sokolov}}, \bibinfo {author}
  {\bibfnamefont {M.}~\bibnamefont {B{\" a}r}}, \bibinfo {author}
  {\bibfnamefont {S.~H.~L.}\ \bibnamefont {Klapp}}, \ and\ \bibinfo {author}
  {\bibfnamefont {I.~S.}\ \bibnamefont {Aranson}},\ }\href {\doibase
  10.1038/s42005-020-0337-z} {\bibfield  {journal} {\bibinfo  {journal}
  {Commun. Phys.}\ }\textbf {\bibinfo {volume} {3}},\ \bibinfo {pages} {76}
  (\bibinfo {year} {2020})}\BibitemShut {NoStop}%
\bibitem [{\citenamefont {Reichhardt}\ and\ \citenamefont
  {Reichhardt}(2021{\natexlab{b}})}]{Reichhardt21}%
  \BibitemOpen
  \bibfield  {author} {\bibinfo {author} {\bibfnamefont {C.}~\bibnamefont
  {Reichhardt}}\ and\ \bibinfo {author} {\bibfnamefont {C.~J.~O.}\ \bibnamefont
  {Reichhardt}},\ }\bibfield  {title} {\enquote {\bibinfo {title} {Active
  matter commensuration and frustration effects on periodic substrates},}\
  }\href {\doibase 10.1103/PhysRevE.103.022602} {\bibfield  {journal} {\bibinfo
   {journal} {Phys. Rev. E}\ }\textbf {\bibinfo {volume} {103}},\ \bibinfo
  {pages} {022602} (\bibinfo {year} {2021}{\natexlab{b}})}\BibitemShut
  {NoStop}%
\bibitem [{\citenamefont {Marchetti}\ and\ \citenamefont
  {Nelson}(1999)}]{Marchetti99}%
  \BibitemOpen
  \bibfield  {author} {\bibinfo {author} {\bibfnamefont {M.~C.}\ \bibnamefont
  {Marchetti}}\ and\ \bibinfo {author} {\bibfnamefont {D.~R.}\ \bibnamefont
  {Nelson}},\ }\bibfield  {title} {\enquote {\bibinfo {title} {Patterned
  geometries and hydrodynamics at the vortex {B}ose glass transition},}\ }\href
  {\doibase 10.1103/PhysRevB.59.13624} {\bibfield  {journal} {\bibinfo
  {journal} {Phys. Rev. B}\ }\textbf {\bibinfo {volume} {59}},\ \bibinfo
  {pages} {13624--13627} (\bibinfo {year} {1999})}\BibitemShut {NoStop}%
\bibitem [{\citenamefont {Banerjee}\ \emph {et~al.}(2003)\citenamefont
  {Banerjee}, \citenamefont {Soibel}, \citenamefont {Myasoedov}, \citenamefont
  {Rappaport}, \citenamefont {Zeldov}, \citenamefont {Menghini}, \citenamefont
  {Fasano}, \citenamefont {de~la Cruz}, \citenamefont {van~der Beek},
  \citenamefont {Konczykowski},\ and\ \citenamefont {Tamegai}}]{Banerjee03}%
  \BibitemOpen
  \bibfield  {author} {\bibinfo {author} {\bibfnamefont {S.~S.}\ \bibnamefont
  {Banerjee}}, \bibinfo {author} {\bibfnamefont {A.}~\bibnamefont {Soibel}},
  \bibinfo {author} {\bibfnamefont {Y.}~\bibnamefont {Myasoedov}}, \bibinfo
  {author} {\bibfnamefont {M.}~\bibnamefont {Rappaport}}, \bibinfo {author}
  {\bibfnamefont {E.}~\bibnamefont {Zeldov}}, \bibinfo {author} {\bibfnamefont
  {M.}~\bibnamefont {Menghini}}, \bibinfo {author} {\bibfnamefont
  {Y.}~\bibnamefont {Fasano}}, \bibinfo {author} {\bibfnamefont
  {F.}~\bibnamefont {de~la Cruz}}, \bibinfo {author} {\bibfnamefont {C.~J.}\
  \bibnamefont {van~der Beek}}, \bibinfo {author} {\bibfnamefont
  {M.}~\bibnamefont {Konczykowski}}, \ and\ \bibinfo {author} {\bibfnamefont
  {T.}~\bibnamefont {Tamegai}},\ }\bibfield  {title} {\enquote {\bibinfo
  {title} {Melting of ``porous'' vortex matter},}\ }\href {\doibase
  10.1103/PhysRevLett.90.087004} {\bibfield  {journal} {\bibinfo  {journal}
  {Phys. Rev. Lett.}\ }\textbf {\bibinfo {volume} {90}},\ \bibinfo {pages}
  {087004} (\bibinfo {year} {2003})}\BibitemShut {NoStop}%
\bibitem [{\citenamefont {Seshadri}\ and\ \citenamefont
  {Westervelt}(1993)}]{Seshadri93}%
  \BibitemOpen
  \bibfield  {author} {\bibinfo {author} {\bibfnamefont {R.}~\bibnamefont
  {Seshadri}}\ and\ \bibinfo {author} {\bibfnamefont {R.~M.}\ \bibnamefont
  {Westervelt}},\ }\bibfield  {title} {\enquote {\bibinfo {title} {Forced shear
  flow of magnetic bubble arrays},}\ }\href {\doibase
  10.1103/PhysRevLett.70.234} {\bibfield  {journal} {\bibinfo  {journal} {Phys.
  Rev. Lett.}\ }\textbf {\bibinfo {volume} {70}},\ \bibinfo {pages} {234--237}
  (\bibinfo {year} {1993})}\BibitemShut {NoStop}%
\bibitem [{\citenamefont {Nagamanasa}\ \emph {et~al.}(2015)\citenamefont
  {Nagamanasa}, \citenamefont {Gokhale}, \citenamefont {Sood},\ and\
  \citenamefont {Ganapathy}}]{Nagamanasa16}%
  \BibitemOpen
  \bibfield  {author} {\bibinfo {author} {\bibfnamefont {K.~H.}\ \bibnamefont
  {Nagamanasa}}, \bibinfo {author} {\bibfnamefont {S.}~\bibnamefont {Gokhale}},
  \bibinfo {author} {\bibfnamefont {A.~K.}\ \bibnamefont {Sood}}, \ and\
  \bibinfo {author} {\bibfnamefont {R.}~\bibnamefont {Ganapathy}},\ }\bibfield
  {title} {\enquote {\bibinfo {title} {Direct measurements of growing amorphous
  order and non-monotonic dynamic correlations in a colloidal glass-former},}\
  }\href {\doibase 10.1038/NPHYS3289} {\bibfield  {journal} {\bibinfo
  {journal} {Nature Phys.}\ }\textbf {\bibinfo {volume} {11}},\ \bibinfo
  {pages} {403--408} (\bibinfo {year} {2015})}\BibitemShut {NoStop}%
\bibitem [{\citenamefont {Reichhardt}\ and\ \citenamefont
  {Reichhardt}(2020{\natexlab{b}})}]{Reichhardt20aa}%
  \BibitemOpen
  \bibfield  {author} {\bibinfo {author} {\bibfnamefont {C.}~\bibnamefont
  {Reichhardt}}\ and\ \bibinfo {author} {\bibfnamefont {C.~J.~O.}\ \bibnamefont
  {Reichhardt}},\ }\bibfield  {title} {\enquote {\bibinfo {title} {Shear
  banding, intermittency, jamming, and dynamic phases for skyrmions in
  inhomogeneous pinning arrays},}\ }\href {\doibase
  10.1103/PhysRevB.101.054423} {\bibfield  {journal} {\bibinfo  {journal}
  {Phys. Rev. B}\ }\textbf {\bibinfo {volume} {101}},\ \bibinfo {pages}
  {054423} (\bibinfo {year} {2020}{\natexlab{b}})}\BibitemShut {NoStop}%
\bibitem [{\citenamefont {Chepizhko}\ and\ \citenamefont
  {Peruani}(2013)}]{Chepizhko13a}%
  \BibitemOpen
  \bibfield  {author} {\bibinfo {author} {\bibfnamefont {O.}~\bibnamefont
  {Chepizhko}}\ and\ \bibinfo {author} {\bibfnamefont {F.}~\bibnamefont
  {Peruani}},\ }\bibfield  {title} {\enquote {\bibinfo {title} {Diffusion,
  subdiffusion, and trapping of active particles in heterogeneous media},}\
  }\href {\doibase 10.1103/PhysRevLett.111.160604} {\bibfield  {journal}
  {\bibinfo  {journal} {Phys. Rev. Lett.}\ }\textbf {\bibinfo {volume} {111}},\
  \bibinfo {pages} {160604} (\bibinfo {year} {2013})}\BibitemShut {NoStop}%
\bibitem [{\citenamefont {Nowak}\ \emph {et~al.}(1998)\citenamefont {Nowak},
  \citenamefont {Knight}, \citenamefont {Ben-Naim}, \citenamefont {Jaeger},\
  and\ \citenamefont {Nagel}}]{Nowak98}%
  \BibitemOpen
  \bibfield  {author} {\bibinfo {author} {\bibfnamefont {E.~R.}\ \bibnamefont
  {Nowak}}, \bibinfo {author} {\bibfnamefont {J.~B.}\ \bibnamefont {Knight}},
  \bibinfo {author} {\bibfnamefont {E.}~\bibnamefont {Ben-Naim}}, \bibinfo
  {author} {\bibfnamefont {H.~M.}\ \bibnamefont {Jaeger}}, \ and\ \bibinfo
  {author} {\bibfnamefont {S.~R.}\ \bibnamefont {Nagel}},\ }\bibfield  {title}
  {\enquote {\bibinfo {title} {Density fluctuations in vibrated granular
  materials},}\ }\href {\doibase 10.1103/PhysRevE.57.1971} {\bibfield
  {journal} {\bibinfo  {journal} {Phys. Rev. E}\ }\textbf {\bibinfo {volume}
  {57}},\ \bibinfo {pages} {1971} (\bibinfo {year} {1998})}\BibitemShut
  {NoStop}%
\bibitem [{\citenamefont {Reichhardt}\ and\ \citenamefont
  {Reichhardt}(2017{\natexlab{b}})}]{Reichhardt17}%
  \BibitemOpen
  \bibfield  {author} {\bibinfo {author} {\bibfnamefont {C.}~\bibnamefont
  {Reichhardt}}\ and\ \bibinfo {author} {\bibfnamefont {C.~J.~Olson}\
  \bibnamefont {Reichhardt}},\ }\bibfield  {title} {\enquote {\bibinfo {title}
  {Depinning and nonequilibrium dynamic phases of particle assemblies driven
  over random and ordered substrates: a review},}\ }\href {\doibase
  10.1088/1361-6633/80/2/026501} {\bibfield  {journal} {\bibinfo  {journal}
  {Rep. Prog. Phys.}\ }\textbf {\bibinfo {volume} {80}},\ \bibinfo {pages}
  {026501} (\bibinfo {year} {2017}{\natexlab{b}})}\BibitemShut {NoStop}%
\bibitem [{\citenamefont {Stenhammar}\ \emph {et~al.}(2015)\citenamefont
  {Stenhammar}, \citenamefont {Wittkowski}, \citenamefont {Marenduzzo},\ and\
  \citenamefont {Cates}}]{Stenhammar15}%
  \BibitemOpen
  \bibfield  {author} {\bibinfo {author} {\bibfnamefont {J.}~\bibnamefont
  {Stenhammar}}, \bibinfo {author} {\bibfnamefont {R.}~\bibnamefont
  {Wittkowski}}, \bibinfo {author} {\bibfnamefont {D.}~\bibnamefont
  {Marenduzzo}}, \ and\ \bibinfo {author} {\bibfnamefont {M.~E.}\ \bibnamefont
  {Cates}},\ }\bibfield  {title} {\enquote {\bibinfo {title} {Activity-induced
  phase separation and self-assembly in mixtures of active and passive
  particles},}\ }\href {\doibase 10.1103/PhysRevLett.114.018301} {\bibfield
  {journal} {\bibinfo  {journal} {Phys. Rev. Lett.}\ }\textbf {\bibinfo
  {volume} {114}},\ \bibinfo {pages} {018301} (\bibinfo {year}
  {2015})}\BibitemShut {NoStop}%
\bibitem [{\citenamefont {Ai}\ \emph {et~al.}(2018)\citenamefont {Ai},
  \citenamefont {Shao},\ and\ \citenamefont {Zhong}}]{Ai18}%
  \BibitemOpen
  \bibfield  {author} {\bibinfo {author} {\bibfnamefont {B.-Q.}\ \bibnamefont
  {Ai}}, \bibinfo {author} {\bibfnamefont {Z.-G.}\ \bibnamefont {Shao}}, \ and\
  \bibinfo {author} {\bibfnamefont {W.-R.}\ \bibnamefont {Zhong}},\ }\bibfield
  {title} {\enquote {\bibinfo {title} {Mixing and demixing of binary mixtures
  of polar chiral active particles},}\ }\href {\doibase 10.1039/c8sm00444g}
  {\bibfield  {journal} {\bibinfo  {journal} {Soft Matter}\ }\textbf {\bibinfo
  {volume} {14}},\ \bibinfo {pages} {4388--4395} (\bibinfo {year}
  {2018})}\BibitemShut {NoStop}%
\bibitem [{\citenamefont {Kolb}\ and\ \citenamefont {Klotsa}(2020)}]{Kolb20}%
  \BibitemOpen
  \bibfield  {author} {\bibinfo {author} {\bibfnamefont {T.}~\bibnamefont
  {Kolb}}\ and\ \bibinfo {author} {\bibfnamefont {D.}~\bibnamefont {Klotsa}},\
  }\bibfield  {title} {\enquote {\bibinfo {title} {Active binary mixtures of
  fast and slow hard spheres},}\ }\href {\doibase 10.1039/C9SM01799B}
  {\bibfield  {journal} {\bibinfo  {journal} {Soft Matter}\ }\textbf {\bibinfo
  {volume} {16}},\ \bibinfo {pages} {1967} (\bibinfo {year}
  {2020})}\BibitemShut {NoStop}%
\bibitem [{\citenamefont {Rodriguez}\ \emph {et~al.}(2020)\citenamefont
  {Rodriguez}, \citenamefont {Alarcon}, \citenamefont {Martinez}, \citenamefont
  {Ram{\' \i}rez},\ and\ \citenamefont {Valeriani}}]{Rodriguez20}%
  \BibitemOpen
  \bibfield  {author} {\bibinfo {author} {\bibfnamefont {D.~R.}\ \bibnamefont
  {Rodriguez}}, \bibinfo {author} {\bibfnamefont {F.}~\bibnamefont {Alarcon}},
  \bibinfo {author} {\bibfnamefont {R.}~\bibnamefont {Martinez}}, \bibinfo
  {author} {\bibfnamefont {J.}~\bibnamefont {Ram{\' \i}rez}}, \ and\ \bibinfo
  {author} {\bibfnamefont {C.}~\bibnamefont {Valeriani}},\ }\bibfield  {title}
  {\enquote {\bibinfo {title} {Phase behaviour and dynamical features of a
  two-dimensional binary mixture of active/passive spherical particles},}\
  }\href {\doibase 10.1039/C9SM01803D} {\bibfield  {journal} {\bibinfo
  {journal} {Soft Matter}\ }\textbf {\bibinfo {volume} {16}},\ \bibinfo {pages}
  {1162} (\bibinfo {year} {2020})}\BibitemShut {NoStop}%
\bibitem [{\citenamefont {Ni}\ \emph {et~al.}(2014)\citenamefont {Ni},
  \citenamefont {Cohen~Stuart}, \citenamefont {Dijkstra},\ and\ \citenamefont
  {Bolhuis}}]{Ni14}%
  \BibitemOpen
  \bibfield  {author} {\bibinfo {author} {\bibfnamefont {R.}~\bibnamefont
  {Ni}}, \bibinfo {author} {\bibfnamefont {M.~A.}\ \bibnamefont
  {Cohen~Stuart}}, \bibinfo {author} {\bibfnamefont {M.}~\bibnamefont
  {Dijkstra}}, \ and\ \bibinfo {author} {\bibfnamefont {P.~G.}\ \bibnamefont
  {Bolhuis}},\ }\bibfield  {title} {\enquote {\bibinfo {title} {Crystallizing
  hard-sphere glasses by doping with active particles},}\ }\href {\doibase
  10.1039/C4SM01015A} {\bibfield  {journal} {\bibinfo  {journal} {Soft Matter}\
  }\textbf {\bibinfo {volume} {10}},\ \bibinfo {pages} {6609} (\bibinfo {year}
  {2014})}\BibitemShut {NoStop}%
\bibitem [{\citenamefont {K{\" u}mmel}\ \emph {et~al.}(2015)\citenamefont {K{\"
  u}mmel}, \citenamefont {Shabestari}, \citenamefont {Lozano}, \citenamefont
  {Volpe},\ and\ \citenamefont {Bechinger}}]{Kummel15}%
  \BibitemOpen
  \bibfield  {author} {\bibinfo {author} {\bibfnamefont {F.}~\bibnamefont {K{\"
  u}mmel}}, \bibinfo {author} {\bibfnamefont {P.}~\bibnamefont {Shabestari}},
  \bibinfo {author} {\bibfnamefont {C.}~\bibnamefont {Lozano}}, \bibinfo
  {author} {\bibfnamefont {G.}~\bibnamefont {Volpe}}, \ and\ \bibinfo {author}
  {\bibfnamefont {C.}~\bibnamefont {Bechinger}},\ }\bibfield  {title} {\enquote
  {\bibinfo {title} {Formation, compression and surface melting of colloidal
  clusters by active particles},}\ }\href {\doibase 10.1039/c5sm00827a}
  {\bibfield  {journal} {\bibinfo  {journal} {Soft Matter}\ }\textbf {\bibinfo
  {volume} {11}},\ \bibinfo {pages} {6187--6191} (\bibinfo {year}
  {2015})}\BibitemShut {NoStop}%
\bibitem [{\citenamefont {Ramananarivo}\ \emph {et~al.}(2019)\citenamefont
  {Ramananarivo}, \citenamefont {Ducrot},\ and\ \citenamefont
  {Palacci}}]{Ramananarivo19}%
  \BibitemOpen
  \bibfield  {author} {\bibinfo {author} {\bibfnamefont {S.}~\bibnamefont
  {Ramananarivo}}, \bibinfo {author} {\bibfnamefont {E.}~\bibnamefont
  {Ducrot}}, \ and\ \bibinfo {author} {\bibfnamefont {J.}~\bibnamefont
  {Palacci}},\ }\bibfield  {title} {\enquote {\bibinfo {title}
  {Activity-controlled annealing of colloidal monolayers},}\ }\href {\doibase
  10.1038/s41467-019-11362-y} {\bibfield  {journal} {\bibinfo  {journal}
  {Nature Commun.}\ }\textbf {\bibinfo {volume} {10}},\ \bibinfo {pages} {3380}
  (\bibinfo {year} {2019})}\BibitemShut {NoStop}%
\bibitem [{\citenamefont {Omar}\ \emph {et~al.}(2019)\citenamefont {Omar},
  \citenamefont {Wu}, \citenamefont {Wang},\ and\ \citenamefont
  {Brady}}]{Omar19}%
  \BibitemOpen
  \bibfield  {author} {\bibinfo {author} {\bibfnamefont {A.~K.}\ \bibnamefont
  {Omar}}, \bibinfo {author} {\bibfnamefont {Y.}~\bibnamefont {Wu}}, \bibinfo
  {author} {\bibfnamefont {Z.~G.}\ \bibnamefont {Wang}}, \ and\ \bibinfo
  {author} {\bibfnamefont {J.~F.}\ \bibnamefont {Brady}},\ }\bibfield  {title}
  {\enquote {\bibinfo {title} {Swimming to stability: Structural and dynamical
  control via active doping},}\ }\href {\doibase 10.1021/acsnano.8b07421}
  {\bibfield  {journal} {\bibinfo  {journal} {ACS Nano}\ }\textbf {\bibinfo
  {volume} {13}},\ \bibinfo {pages} {560} (\bibinfo {year} {2019})}\BibitemShut
  {NoStop}%
\end{thebibliography}%
\end{document}